\documentclass[useAMS,usenatbib,nofootinbib,twocolumn]{aastex6}
\usepackage{amsmath,amsfonts,bbm,graphicx,color}
\usepackage{aas_macros}
\usepackage{url}
\usepackage{multirow}
\usepackage{float}
\usepackage{amsmath}
\usepackage{algorithm}
\usepackage{multirow}
\usepackage[]{algpseudocode}
\makeatletter
\def\BState{\State\hskip-\ALG@thistlm}
\makeatother
\algdef{SE}[SUBALG]{Indent}{EndIndent}{}{\algorithmicend\ }%
\algtext*{Indent}
\algtext*{EndIndent}
\def\be{\begin{equation}}
\def\ee{\end{equation}}
\def\ba{\begin{eqnarray}}
\def\ea{\end{eqnarray}}

\begin{document}
\title{A new approach for obtaining cosmological constraints from Type Ia Supernovae  using Approximate Bayesian Computation}

\author{Elise Jennings \altaffilmark{1,2,4}, Rachel Wolf \altaffilmark{3},  Masao Sako \altaffilmark{3}}

\altaffiltext{1}{Fermi National Accelerator Laboratory MS209, P.O. Box 500, Kirk Rd. \& Pine St., Batavia, IL 60510-0500}
\altaffiltext{2}{Kavli Institute for Cosmological Physics, Enrico Fermi Institute, University of Chicago, Chicago, IL 60637}
\altaffiltext{3}{University of Pennsylvania Department of Physics \& Astronomy, 209 South 33rd Street, Philadelphia, PA 19104-6396}
\altaffiltext{4}{elise@fnal.gov\\}

\begin{abstract}
Cosmological parameter estimation techniques that robustly account for systematic measurement uncertainties will be crucial for the next generation of cosmological surveys. We present a new analysis method, {\it superABC}, for obtaining cosmological constraints from Type Ia supernova (SN Ia)  light curves using Approximate Bayesian Computation (ABC) without any likelihood assumptions. The ABC method works by using a forward model simulation of the data where systematic uncertainties can be simulated and marginalized over. A key feature of the method presented here is the use of two distinct metrics, the `Tripp' and `Light Curve' metrics, which allow us to compare the simulated data to the observed data set without likelihood assumptions. 
The Tripp metric takes as input the parameters of models fit to each light curve with the SALT-II method, whereas the Light Curve metric uses the measured fluxes directly without reference to model fitting.
We apply the {\it superABC} sampler to a simulated data set of $\sim$1000 SNe corresponding to the first season of the Dark Energy Survey Supernova Program (DES-SN).
We investigate the effect of systematic uncertainties
on parameter constraints from the ABC sampler
by including 1\% calibration uncertainties.
Varying five parameters, $\Omega_m, w_0, \alpha$ and $\beta$ and a magnitude offset parameter, with  a CMB prior and no systematics we obtain 
$\Delta(w_0) = w_0^{\rm true} - w_0^{\rm best \, fit} = -0.036\pm0.109$
(a $\sim11$\% 1$\sigma$ uncertainty) using the Tripp metric and 
$\Delta(w_0) =  -0.055\pm0.068$
(a $\sim7$\%  1$\sigma$ uncertainty) using the Light Curve metric. Including calibration uncertainties in four passbands, adding 4 more parameters (9 total), we obtain 
$\Delta(w_0) =  -0.062\pm0.132$
(a $\sim14$\% 1$\sigma$ uncertainty) using the Tripp metric.
Overall we find a $17$\% increase in the uncertainty  on $w_0$ with systematics compared to without. We contrast this with a MCMC approach where systematic effects are approximately included as a fixed uncertainty in the covariance matrix. We find that the MCMC method slightly underestimates the impact of calibration uncertainties for this simulated data set.
\end{abstract}
\keywords{
cosmology: cosmological parameters, dark energy; supernovae: Ia; methods: statistical 
}

\section{Introduction}

Type Ia supernovae (SNe Ia) are a key probe of the dark energy currently driving the late time acceleration of the Universe \citep{1998AJ....116.1009R, 1999ApJ...517..565P}.
Recent cosmological constraints from  e.g.\ the Joint Light-curve Analysis (JLA) collaboration \citep{2014A&A...568A..22B} provide the most stringent constraints to date on both the matter density today, $\Omega_m$, and
the current dark energy equation of state, $w_0$.
However as noted by \citet{2014A&A...568A..22B},  the accuracy of cosmological
constraints obtained using SNe is currently limited by systematic measurement
uncertainties. The next generation of cosmological
surveys are designed to improve the measurement
of $w_0$, and developing parameter estimation techniques which can account for these systematics robustly will be crucial.
In this paper we present a new analysis method for obtaining cosmological constraints from SNe using Approximate Bayesian Computation (ABC).
The ABC method \citep[see e.g.][]{2008arXiv0805.2256B} is a promising alternative to traditional Markov Chain Monte Carlo (MCMC) approaches  and works by using a forward model simulation of the data at every point in parameter space, where systematic uncertainties can be included correctly and marginalized over.
Here the parameter space is $N$ dimensional where $N$ is the number of parameters varied by the sampler.
 A key feature of the analysis method presented here is the use of two distinct metrics that allow us to compare the forward-modeled simulated data to the observed data set without likelihood assumptions. 
 We demonstrate this new sampling method called {\it superABC} by analyzing a simulated data set  based on the  
 first season of the Dark Energy Survey Supernova Program 
(DES-SN) as described in \citet{2015AJ....150..172K}.
 
 Systematic uncertainties may
limit the precision of cosmological parameter constraints from DES-SN and future surveys, and therefore
we need robust methods to account for them. 
Many sources of systematic uncertainty, such as 
sample purity, photometric calibration, selection bias and dust extinction, have been identified in SN Ia analysis studies. 
Other uncertainties exist related to model assumptions in light-curve fitting techniques \citep[see e.g.][]{2011ApJS..192....1C, 2014A&A...568A..22B,2016ApJ...822L..35S} and variations in the SN Ia luminosity with the properties of the host galaxy \citep[e.g.][]{2010MNRAS.406..782S, 2010ApJ...715..743K}. 
A thorough discussion of these effects can be found in e.g. \citet{2011ApJS..192....1C}. 
Recent approaches  have  used simulations, which naturally include systematic uncertainties and correlations, to estimate 
a mean model and covariance matrix  for use in MCMC sampling.
These methods all need to assume some form for the probability of the data given the model and the parameters, which is the likelihood in Bayes's Theorem \citep[e.g.][]{2011ApJS..192....1C, 2014A&A...568A..22B}.
However this approach can result in biased parameter constraints if the assumed likelihood is incorrect or if
the number of simulations used is insufficient to capture the full covariance or if the simulations are run in a fixed
cosmology. 

In summary, current MCMC methods rely on an assumed likelihood for the data, which may be incorrect, and  estimated systematic uncertainties. 
As a result MCMC methods may (i) not be
correctly predicting the impact of these systematics on cosmological parameter constraints or (2) be obtaining incorrect constraints if the assumed likelihood is incorrect.
The ABC method used in this paper allows us to incorporate systematics correctly and does not make any assumption about the likelihood for the data. ABC requires only that one can faithfully simulate the processes that produce the data.
Note that throughout this paper  we use the term `MCMC' to refer to a choice of likelihood, model and method of including systematics which may be estimated or exact in a Bayesian analysis. MCMC is just a sampling technique and there are no shortcomings with this approach when used properly.

DES is carrying out a deep optical and near-infrared survey
of 5000 square degrees of the South Galactic
Cap using the `DECam' 3 deg$^2$ CCD camera \citep{2015AJ....150..150F} 
mounted on the Blanco 4-meter telescope at the
Cerro Tololo Inter-American Observatory (CTIO).
DES-SN consists of a 30 square degree search area (ten 3 square degree fields) in the {\it griz} filter set which is observed roughly once per week.
 Eight are `shallow' fields which 
 are observed to an average depth of 23.5; the other two `deep' fields are observed to an average depth of 24.5.
DES-SN is forecasted to produce a homogeneous sample of a few thousand Type Ia SNe in the redshift
range $0.05<z<1.2$ where spectroscopic observations of the host galaxy will be used to determined the redshift of each SN identified with that host \citep{2012ApJ...753..152B}. The plan is to acquire SN spectra near peak for up to
20\% of this sample and host galaxy spectra for the
 remainder. The remaining 80\% 
for which we get host galaxy spectra will be classified as SNe Ia using the four-band DES photometry \citep{2015MNRAS.452.3047Y}.

For demonstration purposes, we analyze a set of  $\sim1000$ SN Ia assuming an accurate redshift determination obtained by taking a spectrum of the
SN itself or of its host galaxy. A study of the parameter constraints possible with the {\it superABC} sampler when applied
to a simulated photometric sample where 
the redshift comes from the host spectrum without the SN spectrum  is left to future work.

Given the size and complexity of modern cosmological data, Bayesian methods are now standard analysis procedures.
Bayesian inference allows us to efficiently combine datasets
from different probes, to update or incorporate prior information into parameter inference and to carry out model selection or comparison with Bayesian Evidence. 
The standard in cosmological parameter estimation is to adopt a Bayesian approach, where a likelihood function, together with a prior probability distribution function (pdf) for the parameters of interest, are sampled over using an MCMC to simulate from the posterior distribution. There are many public parameter
estimation codes available to the astrophysics community which focus on MCMC methods for analyzing complex cosmological datasets, as well
as calculating the physical analytical models and covariances which are needed in the likelihood \citep[e.g.][]{2002PhRvD..66j3511L, 2014MNRAS.440.1379E,2015A&C....12...45Z}. Evaluating the likelihood for combined probes is a non-trivial task as complex physical data is unlikely to have a simple multi-Gaussian or analytical form. Accounting for modeling  and instrumental systematics, and significant correlations between the parameters of interest and nuisance parameters in either the covariance matrix or likelihood, can be a daunting task \citep{2013PhRvD..88f3537D, 2013JCAP...11..009M}. 
In summary we need to know the likelihood to evaluate the posterior distribution correctly but in many cases we do not.  There are a few ways to deal with this: (i) assume a form for the likelihood (typically Gaussian) and maximize it in a Frequentist analysis to obtain a best fit value rather than the posterior distribution (ii) assume a form for the likelihood and sample from this using a Bayesian MCMC technique (iii) sample from the posterior distribution directly using the ABC likelihood-free method described in this paper.

Cosmology is the latest discipline to employ Approximate Bayesian methods \citep[e.g.][]{2013ApJ...764..116W, 2012MNRAS.425...44C}, a development
driven by both the complexity of the data and covariance matrix estimation, together with the availability of new algorithms for running fast simulations of mock astronomical datasets.
ABC is called   `likelihood free' as 
 explicit evaluation of the likelihood is avoided and replaced with a simulation that produces a mock data set which can be compared to the observed data, while including systematics and correlations self-consistently.

A previous analysis by \citet{2013ApJ...764..116W} applied the ABC technique to SN data from the SDSS-II Supernova Survey
\citep{2014arXiv1401.3317S} and investigated the impact of Type IIP supernovae contamination on the cosmological constraints obtained. The ABC metric employed by \citet{2013ApJ...764..116W}  used the distance modulus
 measured by performing an MLCS2k2 \citep{2007ApJ...659..122J} light-curve fit on the output from the {\rm SNANA} light curve analysis package \citep{2009PASP..121.1028K}.  It is important to note that the distance obtained from the light-curve fit has implicit likelihood assumptions about the data which ideally should be avoided in a complete ABC analysis.
The {\it superABC}  sampler presented in this paper avoids all likelihood assumptions and uses two distinct metrics to compare the simulated and observed data. We fit for both cosmological parameters and SN standardization parameters as well as calibration uncertainties; and we investigate the impact of priors on our constraints.

{\it superABC} makes use of the open source code {\it astroABC} \citep{2016arXiv160807606J}, which is a parallel Python ABC Sequential Monte Carlo (SMC) sampler, for parameter estimation.
 Although in principle any light curve simulation code can be used by {\it superABC}, with the predefined metrics, 
we use {\rm SNANA} and its implementation of the SALT-II light-curve fitter \citep{2010A&A...523A...7G}, as a forward model simulation at every point in parameter space. In this analysis we present cosmological constraints on $\Omega_m$ and $w_0$ with and without accounting for  calibration uncertainties using both uniform priors and priors based on Cosmic Microwave Background (CMB) data.
 These calibration uncertainties can arise from many sources, for example, from image subtraction,
   PSF modeling and nearby bright stars.

In this analysis we are assuming we know the survey conditions and spectral model, as well as the selection effects in the SN Ia dataset.
 We leave an investigation which relaxes these assumptions and includes more systematic uncertainties in our sampling method to future work.
 
 The outline of this paper is as follows: in Section \ref{section:abc_smc} we introduce the ABC method in comparison with the traditional 
MCMC methods and discuss the ABC distance metric and sufficient statistics needed for the analysis.
In Section \ref{sec:usual} we present a brief review of common parameter estimation methods.
In Section \ref{sec:superabc} we present the {\it superABC}  sampler and discuss the forward model simulation as well as the two distance metrics used in detail.
In Section \ref{sec:data} we discuss the simulated data set used in this analysis and  present our results in Section \ref{sec:results}.
We conclude with a summary of our results and a discussion in Sections \ref{sec:summary} and \ref{sec:discussion}.

\section{ Approximate Bayesian Computation}  \label{section:abc_smc}
In Section \ref{sec:b_inf} we give a brief background on Bayesian inference and traditional 
MCMC methods which will be useful when comparing with 
ABC. Readers familiar with these methods can skip to Section \ref{sec:superabc}.
In Section \ref{sec:abc_intro} we describe ABC and motivate its use for cosmological parameter estimation, in Section \ref{sec:abc_algo}
we describe a general ABC Sequential Monte Carlo (SMC) algorithm and in Section \ref{sec:metric} we discuss the ABC distance metric and sufficiency conditions on summary statistics.

\subsection{Bayesian Inference \label{sec:b_inf}}

The fundamental problem in Bayesian statistics is the computation of posterior distributions for the parameters of interest given some data.
We are interested in estimating the posterior pdf for some underlying parameters, ${\bf \theta}$, of a model, $M$, given some data and prior 
information about those parameters. Bayes Theorem allows us to write this posterior distribution in terms of the likelihood for the data, $\mathcal{L}({ D}|M({\bf \theta}))$, and the prior distribution, $\pi({ \theta})$, as
\ba
P({\bf \theta}|{D}) =  \frac{\mathcal{L}({D}|M({\theta})) \pi(\bf{\theta})}{\int \mathcal{L}({{D}}|M({\bf \theta})) \pi(\bf{\theta}) {\rm d}{\theta}}
\ea
where the denominator is referred to as the Bayesian Evidence or marginal likelihood; and the integral runs over all possible parameter values.
The prior probability represents our state of knowledge of the data and may incorporate results from previous datasets; restrict the range for 
physical parameters e.g. masses must be positive; or may be un-informative with little restriction. 
The choice of likelihood for many cosmological analyses is a single or multivariate Gaussian where the mean is evaluated using some physical model and the covariance matrix is measured or estimated either analytically or numerically. 
In this framework the accuracy of the parameter estimation will depend heavily on our choice for the likelihood, as well as the accuracy of the physical model for the data, and how well parameter covariance and correlated systematics are described in the covariance matrix \citep[see e.g.][]{2015A&C....12...45Z,2014MNRAS.440.1379E}.
For a review of probability, parameter inference and  numerical techniques such as MCMC methods please see e.g. \citep{2008ConPh..49...71T, 2009arXiv0906.0664H, Jaynes}.

MCMC techniques are an efficient way to simulate from the posterior pdf when analytical solutions do not exist or are intractable.
An MCMC algorithm constructs a sequence of points in parameter
space, referred to as an MCMC chain, 
which is a discrete time stochastic process where each event in the chain is generated from the
Markov assumption that the probability
of the $(i+ 1)^{th}$ element in the chain only depends on the value of the $i^{th}$
element. 
Markov Chains are called `memory-less' because of this assumption.
A key property of Markov chains is that under certain conditions the distribution of the chain evolves to a
stationary or target state  independently of its initial starting point.
If our target distribution is the posterior pdf then we want the unique
distribution\footnote{The stationary distribution the Markov Chain should asymptote to.} for the Markov Chain to be 
the posterior distribution.
Many MCMC algorithms exist, including the Metropolis-Hastings algorithm \citep{MCMC}, Gibbs sampling, Hamiltonian Monte Carlo, 
importance sampling and ensemble sampling \citep[see e.g.][]{GoodmaneWeare}. 
Each method relies on a proposal distribution (which may have separate parameters which need to be tuned) to advance events in the chain from the starting distribution towards the target pdf.
Once the chain has converged the density of points in the chain is proportional to the posterior pdf.
If the likelihood and model are correct then MCMC will lead to the correct posterior pdf for the model parameters.

\subsection{ABC: parameter inference without likelihood assumptions \label{sec:abc_intro}}

In traditional MCMC approaches the likelihood used (most often a simple multi-Gaussian) is a key assumption in the method. With incomplete analytical expressions for the likelihood or computational restrictions on how accurately we can estimate the covariance matrix, this assumed pdf will be incorrect, leading to biased parameter constraints.
Even if the covariance matrix used is correct we can still obtain incorrect parameter constraints if the assumed form for the likelihood is incorrect.
ABC methods aim to simulate samples directly from the parameter posterior distribution of interest without assuming a particular form for the likelihood.

\subsection{ ABC algorithms \label{sec:abc_algo}}

Given a set of parameters, $\theta$, with associated priors, $\pi(\theta)$ and a forward simulated model for the data vector,
$f(D|\theta)$, 
we can simulate from the posterior distribution, $P(\theta|D)$, by first drawing sample parameters
$\theta^* \sim \pi(\theta)$, 
then simulating a dataset with these parameters 
$D^* \sim f(D|\theta^*)$.
The simplest ABC algorithm is rejection sampling. 
In a  rejection sampling algorithm, we reject $D^*$  unless it exactly equals the true data, $D$.
For discrete data this algorithm would not be practical as many simulated samples would be rejected until a simulation exactly replicates the data.

In practice we make an approximation and accept simulated datasets which are `close' to the true data. This 
notion of simulating a dataset which is close to the observed data
introduces the idea of a distance metric and tolerance level in ABC. 
The distance metric allows us to compare the  data to the simulation and the tolerance level tells us how close the two need to be for us to accept the proposed parameters of the simulation.
We accept proposed parameters $\theta^*$, if 
$\rho(D^* - D) <\epsilon$
where $\rho$ is the distance metric, which could be e.g. the Euclidean norm $||D^* - D||$,  and $\epsilon$ is a tolerance threshold. This procedure produces samples from the pdf
$P(\theta | \rho(D^*-D)<\epsilon)$,
which will be a good approximation of the true posterior if $\epsilon$ is small.
The threshold $\epsilon$ could be chosen to be a constant at each iteration however in practice the algorithm is more efficient if
 $\epsilon$ is initially large, but is decreased
at each step as the distribution converges on the true distribution.

Rather than drawing candidates, $\theta^*$, one at a time, we can
speed up the ABC algorithm by working with large groups or
pools of candidates, called particles, simultaneously. 
At each stage of the algorithm the particles are perturbed and filtered using the distance metric, and eventually
this pool of particles moves closer and closer to simulating from the desired posterior distribution.
This approach is known as Sequential Monte Carlo (SMC) or Particle Monte Carlo sampling and the algorithm is presented in 
Algorithm 
 \ref{alg_1} \citep[see e.g.][]{2008arXiv0805.2256B,2009arXiv0901.1925T, 2010arXiv1001.2058S}.

In 
Algorithm 
 \ref{alg_1} we outline how the particles are filtered and perturbed using a weighted transition kernel. The transition
kernel serves the same purpose as the proposal distribution in a standard
MCMC algorithm. 
The transition kernel, $\mathcal{K}$, specifies the distribution of a random variable that will
be added to each particle to move it around in the parameter space.
Different ABC SMC algorithms can be distinguished
by how sampling weights are assigned to the particles in the pool.
The weighting scheme in ABC SMC  minimizes the
Kullback -- Leibler distance, a measure of the discrepancy between two probability
density functions.  Minimizing the Kullback -- Leibler distance between the desired posterior and
the proposal distribution
maximizes the acceptance probability in the algorithm \citep{2011arXiv1106.6280F}.
For more details on the different choices of kernel as well as optimization techniques \citep[see e.g.][]{2008arXiv0805.2256B, 2011arXiv1106.6280F}.

At iteration $t$, the ABC SMC algorithm proposes parameters from the following
\ba
q_t(\theta) = \begin{cases} \pi(\theta), \qquad \qquad\qquad \qquad \qquad \qquad {\rm if} \, t=0 \\
\sum^{N}_{j=1} w_{j,t-1} \mathcal{K}(\theta_{j,t-1}| \theta_{j,t}, \mathcal{C}_{t-1}) , \qquad  {\rm otherwise}
\end{cases}
\ea
where $w_{j,t-1}$ are the chosen weights for particle $j$ at iteration $t-1$ and $\mathcal{C}_{t-1}$ is the covariance amongst the particles at $t-1$.
This algorithm effectively filters out a particle from the previous weighted pool, $\theta_{t-1}$, then perturbs the result using the kernel ${\mathcal K}$.
We use an optimized Gaussian kernel and set the covariance matrix  to be
twice the weighted covariance matrix amongst all the particles in {\it astroABC} (see Appendix \ref{app:A}) \citep[see][for more details]{2008arXiv0805.2256B, 2011arXiv1106.6280F, 2016arXiv160807606J},
Note also that if the parameters are uncorrelated then a diagonal covariance matrix could be used in the sampler. The details of how the weights are assigned at each iteration is given in Algorithm 
 \ref{alg_1}.

\begin{algorithm}[H]
\caption{ABC SMC algorithm for estimating the posterior distribution for parameters $\theta$ using $N$ particles, the prior distribution $\pi(\theta)$,
given data $D$ and a model for simulating the data $f(D|\theta)$. $\theta_{i,t}$ represents the parameter set for particle $i$ and iteration $t$. Note $\mathcal{N}$ here represents a Normal (Gaussian) distribution.}
\label{alg_1}
\begin{algorithmic}[1]
\BState {Set the tolerance thresholds, $\epsilon_t$ for  $t=0\cdots T$ iterations.}
\Procedure{ ABC SMC LOOP}{} 
\BState{At iteration t=0:}
\For {$1\leq i \leq N$}
\While {$\rho(D,D^*) >\epsilon_0$}
\State Sample $\theta^*$ from prior $\theta^* \sim  \pi(\theta) $
\State Simulate  data  $D^* \sim f(D|\theta^*)$
\State Calculate distance metric $\rho(D,D^*)$ 
\EndWhile
\State Set $\theta_{i,0} \leftarrow \theta^*$
\State Set weights $w_{i,0} \leftarrow 1/N$
\EndFor
\State Set covariance  $\mathcal{C}_{0} \leftarrow  2\mathcal{C}(\theta_{1 :N,0})$
\BState{At iteration $t>0$:}
\For {$1< t < T$}
\For {$1\leq i \leq N$}
\While {$\rho(D,D^*) >\epsilon_t$}
\State Sample $\theta^*$ from previous iteration. $\theta^* \sim \theta_{1:N, t-1} $ with probabilities $w_{1:N,t-1}$
\State Perturb $\theta^*$ by sampling  $\theta^{**} \sim  \mathcal{N}(\theta^*, \mathcal{C}_{t-1})$
\State Simulate  data  $D^* \sim f(D|\theta^{**})$
\State Calculate distance metric $\rho(D,D^*)$ 
\EndWhile
\State Set $\theta_{i,t} \leftarrow \theta^{**}$
\State Set $w_{i,t} \leftarrow \frac{\pi(\theta_{i,t})}{\sum^{N}_{j=1}w_{j,t-1}\mathcal{K}(\theta_{j,t-1}|\theta_{i,t}, \mathcal{C}_{t-1})}$
\EndFor
\State Set covariance $\mathcal{C}_{t}$ using e.g.\ twice weighted empirical covariance \citep{2008arXiv0805.2256B}
\EndFor
\EndProcedure
\end{algorithmic}
\end{algorithm}

\subsection{The ABC metric and sufficient statistics \label{sec:metric}}

Using high-dimensional data can reduce the acceptance rate and reduce the efficiency of the ABC algorithm.
In many cases it may be simpler to work with some lower dimension summary statistic of the data, $S(D)$, e.g. the sample mean,
rather then the full dataset \citep{Marjoram}.
 In this case the chosen statistic needs to be a so-called {\it sufficient statistic}.
A statistic is called a sufficient statistic if 
any information about the parameter of interest which is contained in the data, is also contained in the summary statistic. More formally a statistic $S(D)$ is sufficient for $\theta$, if the distribution $P(D|S(D))$ does not depend on $\theta$.
This requirement ensures that in summarizing the data we have not thrown away constraining information about $\theta$.

The ABC method relies on some distance metric  to
compare the simulated data to the data that were observed.
It is common to use the weighted Euclidean distance,
\ba
\rho(S(D) - S(D^*)) = \left( \sum_i \left( \frac{S(D)_i - S(D^*)_i}{\sigma_i}\right)^2\right)^{1/2}
\ea
 between the  observed and simulated data set or summary statistics as
a metric, where  $\sigma_i$ is the error on the $i^{\rm th}$ summary statistic \citep[see e.g.][]{Beaumont2002}. 
Choosing a summary statistic and distance metric which are sensitive to the parameters of interest is a crucial step in parameter inference.
The success of ABC relies on the fact that if the distance metric is defined by way
of sufficient statistics, then the resulting approximation to the
posterior will be good as long as  $\rho(S(D)-S(D^*))$ is less than some small threshold.
If the model is not able to replicate the data then many samples will be rejected at a given threshold and the ABC algorithm will not converge towards the true posterior distribution.
We outline the two metrics used in the paper in Sections \ref{sec:tripp} and \ref{sec:lc_metric}.

 \section{Review of common parameter estimation methods  }\label{sec:usual}

In this section we discuss two parameter estimation methods which have been used in previous SN analysis studies. The
first approach allows us to
 obtain cosmological parameter constraints
from a  SN dataset by using the fitted SALT-II parameters of
epoch of peak brightness ($t_0$), amplitude ($x_0$), stretch ($x_1$), and color ($c \sim B -V$ at  $t_0$),
for each event. 
A distance modulus is determined for each event by standardizing the SN brightness and using the Tripp relation \citep{1998A&A...331..815T},
\ba
\mu= m_B + \alpha x_1 - \beta c - M_0
\ea
where $\mu$ is the distance modulus,
$m_B = -2.5{\rm log} \,x_0$, $M_0$ is the rest-frame magnitude for a SN Ia with $x_1 = c=0$ and $\alpha, \beta$
are global parameters used to standardize the SN Ia brightness. This procedure 
for obtaining distances using the Tripp relation
is implemented in the {SALT2mu } program \citep{2011ApJ...740...72M}. We shall refer to $\alpha$ and $\beta$ as the supernova standardization
parameters in this paper. In SALT2mu the free parameters $ \alpha$ and $\beta$ are determined in a {\it fixed fiducial} cosmology using a maximum likelihood
estimation and the 
distance modulus for each event is then obtained assuming the Tripp relation. After light-curve fitting, cosmological parameter constraints 
are then obtained using either  a frequentist approach
 maximizing a Normal (Gaussian) likelihood pdf, or a Bayesian approach sampling over the product of this likelihood with the prior to simulate samples from
the posterior distribution. A simple example of the  likelihood assumed is 
\ba
&\mathcal{L}(\mu | \mu_{model}(z,\Omega_m,w_0,\cdots)) \propto \\ 
&\exp\{-\frac{1}{2}\sum_i \left( \frac{\mu^i - \mu_{\rm model}(z^i, \Omega_m,w_0)}{\sigma^i}\right)^2\} \nonumber
\label{eq:like}
\ea
where $\mu^{i}$ is the distance modulus for an individual SN event (assuming that each is independent and drawn from a Normal distribution), with associated error $\sigma^i$; and in a flat universe $(\Omega_k=0)$,
\ba
\mu_{\rm model}(z^i,\Omega_m,w_0,\cdots)  &=&  5 {\rm log}_{10} \frac{c(1+z)}{h}\int_0^{z^i} {\rm d}z' \frac{1}{E(z')} \nonumber \\ 
\label{eq:mu}
\ea
where $h$ is the Hubble parameter and 
\ba
E(z) = \sqrt{\Omega_m (1+z)^3 + (1 - \Omega_m)e^{3 \int_0^z {\rm d}\ln(1+z')[1+w(z')]} } \, . \nonumber \\ 
\ea
Here $w$ is the dark energy equation of state with present value $w_0$. 
The associated error, $\sigma_i$, includes contributions from  sample variance, correlations between $c, x_1$ and $m_B$, intrinsic scatter, redshift
uncertainty, peculiar velocity uncertainty and lensing uncertainty.

Note in the {\it superABC} sampler we will fit for cosmological parameters of interest and the supernova standardization
parameters $\alpha$ and $\beta$ {\it simultaneously}. Ultimately this makes the ABC approach presented here
very different to the  two stage fitting technique presented above. 
As an example consider  fitting for the following parameters $\Omega_m, w_0, \alpha$ and $\beta$ given some
data $D$. Using the ABC sampling technique on this example, we
are simulating samples from a 4 dimensional posterior pdf $P(\Omega_m, w_0, \alpha,\beta |D)$.
In the commonly used two stage fitting technique the constraints on $\alpha$ and $\beta$ are  obtained from a {\it conditional} pdf, $P(\alpha, \beta |D, \Omega_m, w_0)$,
and, following this, the constraints on $\Omega_m, w_0$ are simulated from a conditional pdf $P(\Omega_m, w_0,| \alpha, \beta,D)$.
Ultimately this means that the two methods are very different and we should not expect them to yield equivalent results.
However, the results may be similar in specific cases e.g.\ where the first fit for $\alpha$ and $\beta$ gives unbiased results and informative priors on used in the subsequent sampling step.

The fitted parameters output from SALT-II fitting program, ($t_0, x_0, c,$ and $x_1$), 
are a biased sample from an intrinsic parent distribution of color and stretch.
This bias occurs in sample selection and light-curve fitting to the data which has both intrinsic scatter and noise.
This bias has been examined in recent works \citep{2011ApJS..192....1C, 2014A&A...568A..22B, 2016ApJ...822L..35S}. 
In typical likelihood-based analyses, the likelihood assumes a complete set of SNe Ia. Therefore, working with biased samples will result in biased estimates of the model parameters. 
One of the key advantages of the ABC approach is that as long as our simulated dataset is a correct forward model simulation of the data, then,
when creating a sufficient summary statistic of each (Section \ref{sec:metric}), these biases are naturally taken into account. For example the same selection effects and light-curve fitting method are applied to both the simulation and the data.

The second parameter estimation method which we discuss uses a MCMC sampler with the likelihood adopted in \citet{2014A&A...568A..22B}.
In this method the likelihood function  assumed  is different from Eq.\ \ref{eq:like} in that both the `data' and the model depend on the parameters (equation 15 in \citet{2014A&A...568A..22B}). In this approach cosmological parameters, $\Omega_m$ and $w_0$ are fit simultaneously with $\alpha$ and
$\beta$. 
In the following sections we present 
some results using a likelihood module based on equation 15 of \citet{2014A&A...568A..22B} and a MCMC sampler in the publicly available parameter estimation code CosmoSIS \citep{2015A&C....12...45Z}. 
The effect of calibration uncertainties are estimated
using the method described in equations 5 and 6 
of
\citet{2011ApJS..192....1C}.  
In this method calibration uncertainties are estimated and incorporated as a fixed uncertainty in the covariance matrix and do not vary from point to point in parameter space. Note that both $\alpha$ and $\beta$ are allowed to vary in the covariance matrix at every point in parameter space.

In this analysis we present results accounting for calibration uncertainties in the four DES passbands in both the {\it superABC} sampler  and using an assumed Gaussian likelihood in a MCMC sampler from CosmoSIS when wide uniform and CMB priors are employed in each. 
Note we present these two results to contrast the two different methods of incorporating systematics in each. 
The main advantage of ABC is that the systematic effects are implemented consistently at every point in parameter space and if there are any correlations with other parameters they are marginalized over correctly. 
To  account for systematic effects due to calibration uncertainties in the {\it superABC} sampler, we include four extra parameters, $\mathcal{Z}_p^{\it g}, \mathcal{Z}_p^{\it r} , \mathcal{Z}_p^{\it i}, \mathcal{Z}_p^{\it z}$, each with a Gaussian prior
$\mathcal{Z}_p^k \sim \mathcal{N}(0,0.01)$ where $k$ represents one of  the four passbands. This is
implemented in the forward model simulation by changing the observed zero point, which is a mechanism for simulating calibration uncertainties.

With an assumed likelihood in an MCMC sampler calibration systematics are incorporated 
approximately as a fixed uncertainty in the covariance matrix and so it is not clear if the effect of systematics has been overestimated or underestimated .
It is important to distinguish between a fundamental limitation of incorporating systematics versus a poor implementation  of them in
 the likelihood. With the correct analytical expression or estimate from simulations systematics can be incorporated exactly in the likelihood.
 However, if the systematic uncertainties are degenerate with the cosmological parameters in a way which varies
  from point to point in parameter space, e.g. changing calibration uncertainties are degenerate with changing $w_0$, then using a fixed covariance matrix in the likelihood can either over or underestimate the effect of these systematics.

Correctly accounting for calibration systematics, the  {\it superABC} sampler is fitting for nine parameters simultaneously 
while with estimated calibration systematics the  MCMC sampler is fitting for five parameters.  
Without systematics, both samplers are varying five parameters and although the ABC and MCMC methods are very different we would expect the results to agree in this case if the Gaussian likelihood assumed in MCMC is correct.

In previous studies using MCMC techniques the effect of many different systematic uncertainties such as core collapse contamination or host-galaxy mis-match has either been neglected or approximately accounted for using corrections from simulations in a fixed cosmology.
In the ABC method we can correctly incorporate the effect of these systematic uncertainties once these are included in the forward model simulation.
This is the real advantage of ABC over MCMC and 
once systematics are correctly accounted for
 we
do not necessarily expect smaller parameter uncertainties from ABC compared to MCMC. 

 \section{ The {\it \MakeLowercase{super}ABC} sampler }\label{sec:superabc}

 In Section \ref{subsec:superabc} we present the supernova specific ABC sampling code, {\it superABC}, and we discuss the forward model simulation which is used in the sampler.
  In Section \ref{sec:dist_metrics} we present the two distance metrics used in this analysis.
  In this paper {\it superABC} is a specific implementation of ABC with the novel metrics we present in Section \ref{sec:dist_metrics}.
 \newline
 \vspace{1.cm}
 \subsection{Overview \label{subsec:superabc}}

The open source code {\it superABC} is will be made publicly available at https://github.com/EliseJ/superabc. 
  The  {\it superABC} code uses {\it astroABC} \citep{2016arXiv160807606J}, which is a parallel Python ABC SMC sampler, for parameter estimation.
In principle any light curve simulation code could be used in {\it superABC}, with the predefined metrics.
In this paper we use  the {\rm SNANA} light curve analysis package \citep{2009PASP..121.1028K} and its implementation of the SALT-II light-curve fitter \citep{2010A&A...523A...7G}, as a forward model simulation at every point in parameter space.  Note that ideally the simulation package  used should be able to 
produce a realistic sample of SN light curves (flux and uncertainties) in order to obtain accurate constraints on the cosmological parameters.
We are assuming that the forward model simulation is able to accurately simulate real data and leave an investigation of potential mismatches between the true SN parent distribution and the model distribution to future work.
 
There are several features of {\it superABC} which are designed to optimize the sampling procedure and these are presented in Appendix \ref{app:D}.
The {\it superABC} sampler comes with the choice of two distance metrics which are described in detail in 
Section \ref{sec:dist_metrics} and there are end-user options for new distance metrics to be defined.

One of the main advantages of using simulations in the ABC sampling technique is the fact that systematic uncertainties, which are not easily incorporated analytically into
a likelihood function, are correctly included and marginalized over during sampling.
In the following we assume  that our simulation correctly generates measurement noise, SN Ia intrinsic scatter, and selection biases.
Note that in this work the forward model simulation code used in 
  {\it superABC} is the same as that which was used to generate the data sample described in Section \ref{sec:data}. As this simulation is capable of producing a realistic sample of SN light curves
  we do not expect this choice to influence the results if applied to a real data set.
As any forward model simulation can be used in the {\it superABC} sampler it would be interesting to follow up this analysis using e.g.\ two different simulation packages or using a  different selection function in the forward model simulation to the one used to generate the mock data.

 \subsection{Distance metrics in {\it superABC}  \label{sec:dist_metrics}}

 In ABC sampling the distance metric is needed to 
compare the simulated samples to the observed data at every point in parameter space.
In the following sections we represent parameters which are being varied by the sampler using a star superscript e.g. $\Omega_m^*$.
 In the {\it superABC} sampler we consider two different distance metrics independently. The first metric is based on the Tripp relation \citep{1998A&A...331..815T} and is 
 described in Section \ref{sec:tripp}; we shall refer to this as the `Tripp' metric.  The second metric compares the data and simulated  light curve fluxes at every proposed point in parameter space, and does not use the SALT-II formalism. 
  We shall refer to this metric as the `Light Curve' metric and it is described in Section \ref{sec:lc_metric}.
  Note that in practice determining whether or not a summary statistic is `sufficient' (Section \ref{sec:metric}) amounts to testing if the true posterior distribution for the parameters are recovered correctly after ABC sampling on the data. If a summary statistic in a metric is not able to constrain a parameter at all then it is certainly not sufficient for that parameter.

 \subsubsection{The `Tripp' metric \label{sec:tripp}}

The first metric we consider is based on the Tripp relation
and uses the SALT-II  fitted light curve parameters for each SN. 
The Tripp metric is the absolute difference between two
weighted Euclidean distances given by
\ba
\label{eq:tripp2}
\Delta_{\rm data} = \\ \nonumber
\frac{1}{N_{\rm data}}\sum_i^{N_{\rm data}} & \frac{[\mu(z_i^{data}, \theta^*) - (m^{data} _{b,i} + \alpha^*x^{data}_{1,i}- \beta^*c^{data} _{i} - M_0 -\delta M_0^* )]^2}{ \sigma_{m_{b,i}}^2 + (\alpha^*\sigma_{x_{1,i}} )^2 + (\beta^*\sigma_{c_i})^2  + \sigma_{\rm int}^2}\\
\label{eq:tripp}
\Delta_{\rm sim} = \\ \nonumber
\frac{1}{N_{\rm sim}}\sum_j^{N_{\rm sim}} & \frac{ [\mu(z_j^{sim}, \theta^*) - (m^{sim} _{b,j} + \alpha^*x^{sim}_{1,j}- \beta^*c^{sim} _{j} - M_0 - \delta M_0^* )]^2}{ \sigma_{m_{b,j}}^2 + (\alpha^*\sigma_{x_{1,j} })^2 + (\beta^*\sigma_{c_j})^2  + \sigma_{\rm int}^2}\, ,
\ea
where $\mu(z_i^{data}, \theta^*)$ is the distance modulus evaluated in the proposed ABC cosmology, $\theta^* = (\Omega_m^*, w_0^*)$,   but  at the 
{\it measured} redshift of the data event. The variables $ \alpha^*, \beta^*$ and $\delta M_0^*$ are the stretch and color standardization parameters, and the magnitude offset parameter respectively. The variables  $x_{1,i}^{data}$ and  $c_i^{data}$ are the SALT-II {\it fitted} stretch and color 
parameters and $m_{b,i}^{data}$ is the magnitude (see Section \ref{sec:usual}) for each data event, $i$. The associated $1\sigma$ errors  are $\sigma_{m_{b,i}}, \sigma_{x_{1,i}} $ and $\sigma_{c_i}$. $\sigma_{\rm int}$ is the error due to intrinsic scatter which is fixed to a value of 0.11.
$M_0$ is fixed to a value of $-19.35$. In Eqs. \ref{eq:tripp2} and \ref{eq:tripp} above $N_{\rm data}$ and $N_{\rm sim}$ are the number of data and simulated events respectively.
The parameters are similarly defined  for  $\Delta_{\rm sim}$ for the simulated sample. Note that the distance modulus is evaluated in the proposed ABC cosmology at the redshift of the simulated light curve.
The Tripp metric for the {\it superABC} sampler is 
\ba
\rho &=& |\Delta_{\rm data} - \Delta_{\rm sim}| \, ,
\label{eq:rho}
\ea
 and is evaluated at every point in parameter space during sampling; see Section  \ref{sec:metric} for the general definition of the ABC metric.

In the sampler we simulate samples with approximately the same number of events as the data at every point in parameter space ($\sim$ 1000). However due to selection cuts being applied to a random selection of SN properties, the number of simulated events is not necessarily exactly equal to the number of events in the data, $N_{\rm data} \ne N_{\rm sim}$ in Eqs. \ref{eq:tripp2} and \ref{eq:tripp}.
In principle this parameter could also be varied in the {\it superABC} sampler if the forward model simulation includes $\sigma_{\rm int}$ as in input variable.
 
 \subsubsection{The `Light Curve' metric \label{sec:lc_metric}}

 The second metric we consider in the {\it superABC} sampler compares the data and simulated  light curve fluxes output from {\sc SNANA} for each observation, for each event (each SN), at every proposed point in the $N$ dimensional parameter space, where $N$ is the number of parameters varied by the sampler. 
  
{\it  \bf Overview of the metric \label{subsec:overview}}

To motivate this metric we begin by considering the case where the cosmological parameters are exactly the same in the simulation and the data.
During an ABC sampling run, for each data event, simulated light curve parameters are generated within $2\sigma$
of the color, stretch,  date of estimated peak luminosity in $g$ band, and redshift\footnote{The variance of $2\sigma$ is used here so that the simulated event in a four dimensional parameter space, ($x_1,c,t_0,z$),
 is sufficiently close to the data event without inefficiently simulating events until we have an exact match to the data.}.
Each observed light curve is paired with a simulated one based on the light curve properties of the data event.

Each data event represents a random draw of color and stretch drawn from an intrinsic distribution  i.e.
the true distribution of colors and stretches. Here we are assuming we know the intrinsic populations of SN Ia color and stretch that correlate with luminosity,\footnote{In practice we fix the parameters which describe the intrinsic populations. In principle these do not have to be fixed and can vary in the ABC sampler.} and so the only difference between our simulated events and the data will be due to
statistical fluctuations, as each is a different realization
from the same distribution.

Therefore  a single data and simulated  light curve pair will differ from one another due to statistical fluctuations (different draws from the intrinsic population) in color, stretch and intrinsic scatter even though the cosmological parameters are exactly the same. 
As our primary interest is in constraining cosmological parameters any ABC metric must be sensitive to changes in e.g.\ $\Omega_m$ and $w_0$. It must account for these statistical fluctuations and not mistakenly associate a data-simulated light curve mismatch as being due to differences in cosmology. 
In order to do this, prior to  sampling,  we create a  `reference  difference' probability distribution function (with fixed assumptions that we know the redshifts and the populations of SN Ia color and stretch that correlate with luminosity of our data).

We shall describe the  method for generating the reference difference distribution in detail below. 
We also give the details of how this metric works in practice.

\vspace{1cm}

{\it  \bf The reference difference pdf \label{subsec:ref_sim_data_pdf}}

Prior to any sampling,  we create a  `reference  difference' probability distribution function with the following steps: 

\begin{itemize}
\item A {\it mock} data set, in a fixed known cosmology, which has the same sample size as the data, is created using the forward model simulation.
\item The same number of  light curve events as the mock data are simulated  {\it in the same fixed cosmology}.  Note that we use different random seeds for the mock data and the simulated sample here.
\item We define a random variable $\delta^{cc} $ as
\ba
\delta_i^{cc} &\equiv& \frac{d_i^{cc} - s_i^{cc}}{\sigma(d_i^{cc})}
\label{eq:diff_lc}
\ea
where $d_i$ and $s_i$ are the fluxes at a single epoch in the mock data and simulated light curves in a matched pair respectively and $\sigma(d_i)$ is the error on the data light curve flux. Here the superscript $^{cc}$ means that the mock data and simulation both have the same cosmology.
Note $\delta_i^{cc}$ is a random variable as it is a combination of the mock data and the simulation sample which are both considered random variables in any Bayesian analysis.
\item We bin  $\delta_i^{cc}$ into $N_{\rm bins}$ bins of width $B_{\rm width}$  and normalize the resulting distribution. 
As a very simple example: for a sample of 500 SN, each observed 5 times, the histogram would be the distribution of 2500 differences in flux.
\item The normalized distribution of $\delta^{cc}$ represents the `reference  difference' probability distribution and is 
saved for use in the {\it superABC} sampler.  Note this is one single reference difference pdf which characterizes the statistical fluctuations between the mock data and the simulated sample.
\end{itemize}

We  refer to this saved distribution as  the {\it reference difference} pdf, $p(\delta^{cc})$. It represents the {\bf expected deviations in the light curves if the cosmology is the same}, as well as  the redshift and the  SN Ia color and stretch population assumptions in both the mock data and the simulated sample.  Note the distribution of the random variable $\delta^{cc}$ is similar to a Gaussian with mean 0, although it is more sharply peaked with wider tails.
In this paper we create $p(\delta^{cc})$ using three fixed cosmologies i.e.\ all of the steps above are followed for three different cosmologies ($\Omega_m$ = 0.23, 0.3 and 0.46 and $w_0 = -1$). The reference  difference pdf is the average from these three. We have found in practice that using more than three cosmological models  or varying more parameters has little impact on the reference difference pdf.
We have tested that $p(\delta^{cc})$ is very similar, independent of the cosmology. Just as long as the simulation and mock data have the same cosmology, the distribution of the random variable, $\delta_i^{cc}$, is approximately constant. There are small differences between the three pdfs but the key point is that 
these are much smaller then the difference between a data and simulated sample set which have different cosmological parameters.  In Appendix \ref{app:A} we show the normalized reference  difference pdfs in these three cosmologies.

{\it  \bf Using the Light Curve metric in the sampler \label{sec:lc_metric_sampling}}

For the data sample described in Section \ref{sec:data} (note this is distinct from the mock data in Section \ref{subsec:ref_sim_data_pdf} which we use to make the reference pdf), we run the  {\it superABC} sampler which, for every particle, proposes a trial set of cosmological parameters, $\theta^*$, representing a trial cosmology, $\tilde{c}$.
We distinguish between these two as $\theta^*$ represents only the parameters which are varying in the sampler while $\tilde{c}$ represent all parameters
needed to specify the cosmological model. For example, we may choose to vary only $\theta^* = \{\Omega_m\}$ in the sampler but the cosmological model is specified by the 6 parameters of the $\Lambda$CDM model.
The data sample is described by an unknown `true' cosmology which we denote as $c_T$.

Given $\theta^*$ in cosmology $\tilde{c}$ we generate approximately the same number of  simulated light curve events as the data
and evaluate the following random variable for each epoch in the  data light curve,

\ba
\delta_i^{c_T\tilde{c}} \equiv \frac{d_i^{c_T} - s_i^{\tilde{c}}}{\sigma(d_i^{c_T})} \, .
\ea

We bin the observed frequencies of $\delta^{c_T\tilde{c}}$ into $N_{\rm bins}$ bins of width $B_{\rm width}$ and denote this  {\it unnormalized} distribution as $\mathcal{O}_{{c_T \tilde{c}}}$ where the number of $\delta_i^{c_T\tilde{c}}$ is $N_{obs}$\footnote{$s_i$ is the simulated flux at each observation in the data light curve. This is done in practice by creating a spline to the simulated light curve, after 
scaling the simulated fluxes to the redshift of the data light curve, using the squared ratio of the luminosity distances at each redshift  in the simulated cosmology. The spline is then evaluated at each observation in the data light curve.}. This is the observed distribution of differences.

If our trial cosmology is correct i.e. if $c_T = \tilde{c}$, then we would expect $\mathcal{O}_{{c_T \tilde{c}}}$ to be drawn from the reference difference probability distribution function, $p(\delta^{cc})$ (Section \ref{subsec:ref_sim_data_pdf}).
Given the  reference difference pdf, the {\it expected} distribution of differences for the data sample is

\ba
\mathcal{E}_{cc} = p(\delta^{cc}) N_{\rm obs} B_{\rm width}\,. 
\label{eq:e_cc}
\ea

We use the same number of bins as the reference metric, $N_{\rm bins}$ to find $\mathcal{E}_{cc}$. 
Since $p(\delta^{cc})$ is normalized integrating Eq. \ref{eq:e_cc} over all bins gives the total number observed, $N_{\rm obs}$, as expected.

Finally, the Light Curve metric is defined as the Pearson's chi-square test statistic given observed and expected frequencies:

\ba
\label{eq:chi2}
\chi^2 &=& \sum_{j=0}^{N_{bins}}  \chi^2_j \, ,\qquad  {\rm where} \\ \label{eq:lc_metric}
\chi^2_j &\equiv& \frac{(\mathcal{O}_{{c_T \tilde{c}},j} - \mathcal{E}_{cc,j} )^2}{\mathcal{E}_{cc,j} }\, .  \label{eq:lc_metric2} 
\ea

In the {\it superABC} sampler the metric is $\rho = \chi^2$.
The nice feature of this metric is that the test statistic is distributed according to the $\chi^2$ distribution with $N_{\rm bins}$ degrees of freedom and we present some illustrative 
examples of this in Appendix \ref{app:A}.

  \section{ The data set and SNANA simulation}\label{sec:data} 
 
 Throughout this paper we use a  simulated data set constructed from SNANA simulations based on the   first DES Supernova program season  \citep{2012ApJ...753..152B}. We refer to this as our data in the 
 rest of the paper to distinguish it from the simulated outputs from the {\it superABC} sampler.
 The data set consists of 1070 light curves in the redshift range $0.01<z<1.2$ based on the cadence and observing conditions from the DES  supernova program \citep{2015AJ....150..172K, 2015AJ....150..150F} in the {\it griz} filter set.
  
  For each SN event the {\sc SNANA} simulation code generates  a realistic flux and uncertainty for each  observation. These fluxes are  translated 
  into simulated observed fluxes and uncertainties using a survey specific library \citep[see e.g. Fig. 1 in][for an example excerpt from a such a library for the SDSS-II SN Survey]{2009PASP..121.1028K}.
 For each DES supernova observation, the simulated magnitude is converted into a flux using the image zero point and CCD gain. The simulated 
flux uncertainty is computed from the point-spread function (PSF), sky noise, and zero point.

 \begin{figure}
\begin{center}
\includegraphics[height=2.5in,width=3.in]{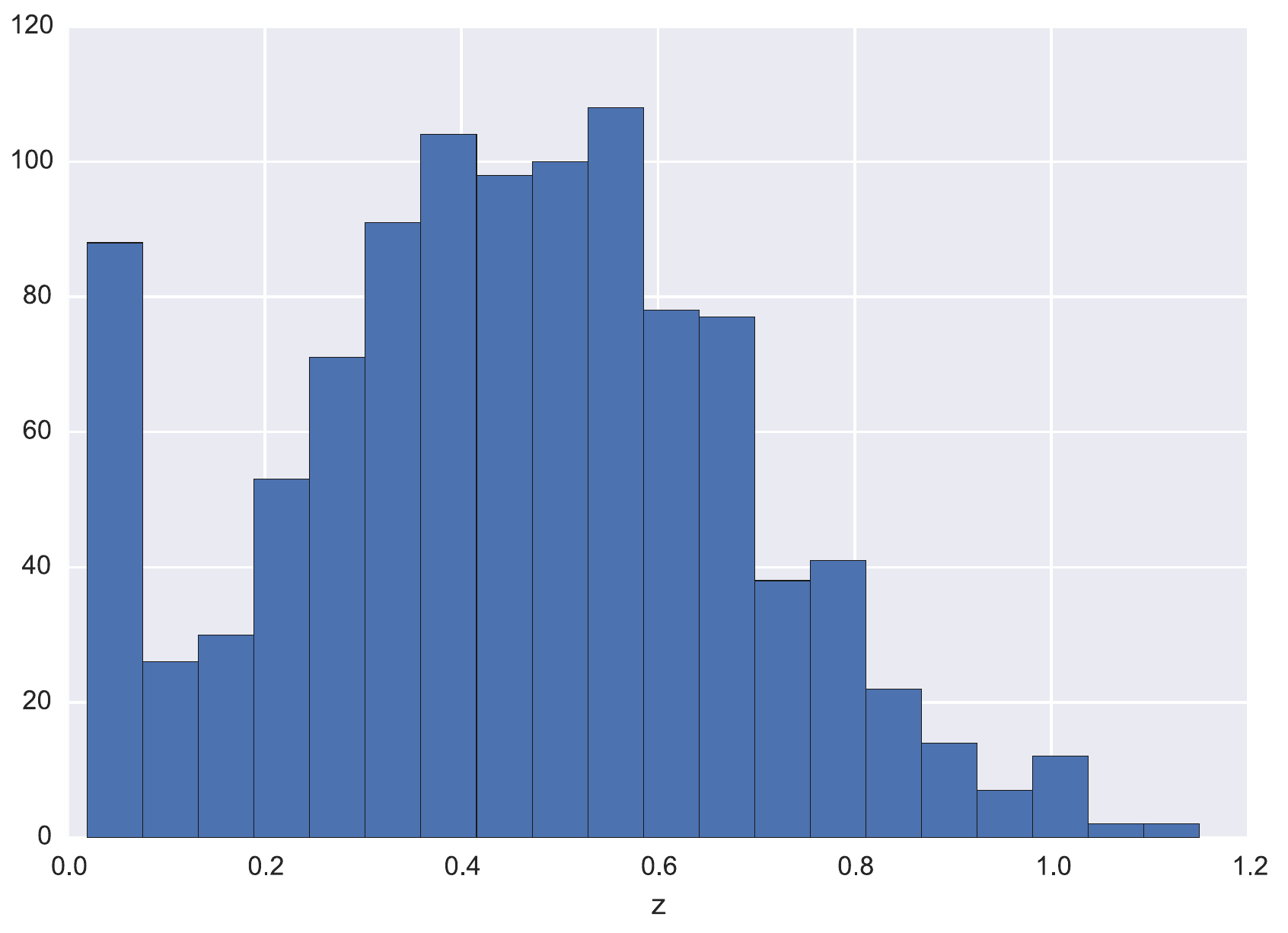}
\caption{The redshift distribution of the data set used in this paper. A low-redshift sample ($0.01<z < 0.08$) makes up
$\sim$9\% of the total distribution. }
\label{fig:z_dist}
\end{center}
\end{figure}
 
  The supernova model magnitudes are generated from the {\sc SALT-II} light curve model ``G10" in which 70\% of the contribution to the Hubble 
 residuals is from achromatic variation and  30\% from chromatic variation.
   The redshift distribution was generated with the redshift-dependent volumetric rate, $R$,
taken from \citet{2008ApJ...682..262D}, with $R(z) \propto (1+z)^{1.5}$.
An artificial low redshift sample ($0.01<z < 0.08$) is generated
with the same passbands and depth as for the DES sample and comprises $\sim$9\% of the total distribution.
In this analysis  we ignore
contamination from core collapse SNe that would occur in a photometric 
analysis, and simulate only spectroscopically confirmed SN Ia.
 The redshift distribution of our data set is shown in Fig.\ \ref{fig:z_dist}.
 Our simulated data set was generated with the following cosmology and standardization parameters:
 $\Omega_m = 0.3, \Omega_{\Lambda} = 0.7, h=0.7, w_0=-1, \alpha=0.14$ and $\beta=3.2$. The simulation and fitting files used to produce the simulated dataset in this paper are available online at  https://github.com/EliseJ/superabc.

\vspace{1cm}
\section{Results} \label{sec:results}

In Section \ref{sec:results_tripp} we present the results from the {\it superABC sampler} using the Tripp metric, described in Section \ref{sec:tripp},
both with and without systematic uncertainties.
In Section \ref{sec:results_lc}  we present the results from using the Light Curve metric discussed in Section \ref{sec:lc_metric}.

To  account for systematic effects due to calibration uncertainties, we include four extra parameters, $\mathcal{Z}_p^{\it g}, \mathcal{Z}_p^{\it r} , \mathcal{Z}_p^{\it i}, \mathcal{Z}_p^{\it z}$,  which are sampled over in the ABC sampler, using the following Gaussian prior
$\mathcal{Z}_p^k \sim \mathcal{N}(0,0.01)$ where $k$ represents one of  the four passbands. This is
implemented in the SNANA simulation by changing the observed zero point, which is a mechanism for simulating calibration uncertainties.

Throughout we use the {\it superABC} sampler with an adaptive tolerance threshold based on the 75$^{\rm th}$ quartile of the distances ($\rho$ in Algorithm \ref{alg_1}) in the previous iteration and a weighted covariance matrix estimation in order to perturb the particles at each iteration \citep[see][for more details on these settings]{2016arXiv160807606J}.
The tolerance level, $\epsilon$, is a constant which decreases at each iteration to ensure that the simulated sample matches the true data set more closely and we can recover the correct posterior distribution.
In all runs we remove $\sim 20$\% of the steps as burn-in and all errors reported are using the resulting steps.
We report wall clock times for running the {\it superABC} sampler in Appendix \ref{app:B}.

\subsection{Priors}\label{sec:priors}

Where stated, we have used a CMB prior on the cosmological 
parameters $\Omega_m$ and $w_0$. In practice we do this using the publicly available MCMC chains ($TT + {\rm low}P$, allowing the dark energy equation of state parameter, $w_0$, to vary) from the Planck Collaboration \citep{2015arXiv150201582P}.
We use the marginal 1D pdfs for both parameters $P(\Omega_m)$ and $P(w_0)$ from these chains as priors in the {\it superABC} sampler using inverse transform sampling. The mean and standard deviations of these 1D pdfs are
$\Omega_m = 0.338\pm0.099$ and $w_0=-1.032\pm0.475$. Note using the marginal pdfs is a conservative choice here as all possible correlations with the other parameters are included and marginalized over which will inflate our constraints.

When not using Planck priors for $\Omega_m$ and $w_0$ we use wide uniform priors for each:
$\Omega_m \sim \mathcal{U}(0.05,0.95)$ and $w_0 \sim \mathcal{U}(-2.5,-0.2)$.
The priors used for the remaining parameters are:
$\alpha \sim \mathcal{U}(0.05,0.25)$, $\beta \sim \mathcal{U}(1.0,5.0)$ and $\delta M_0 \sim  \mathcal{N}(0,0.02)$.

\subsection{Constraints using the Tripp metric}\label{sec:results_tripp}

\begin{figure}
\begin{center}
\includegraphics[height=3in,width=3.5in]{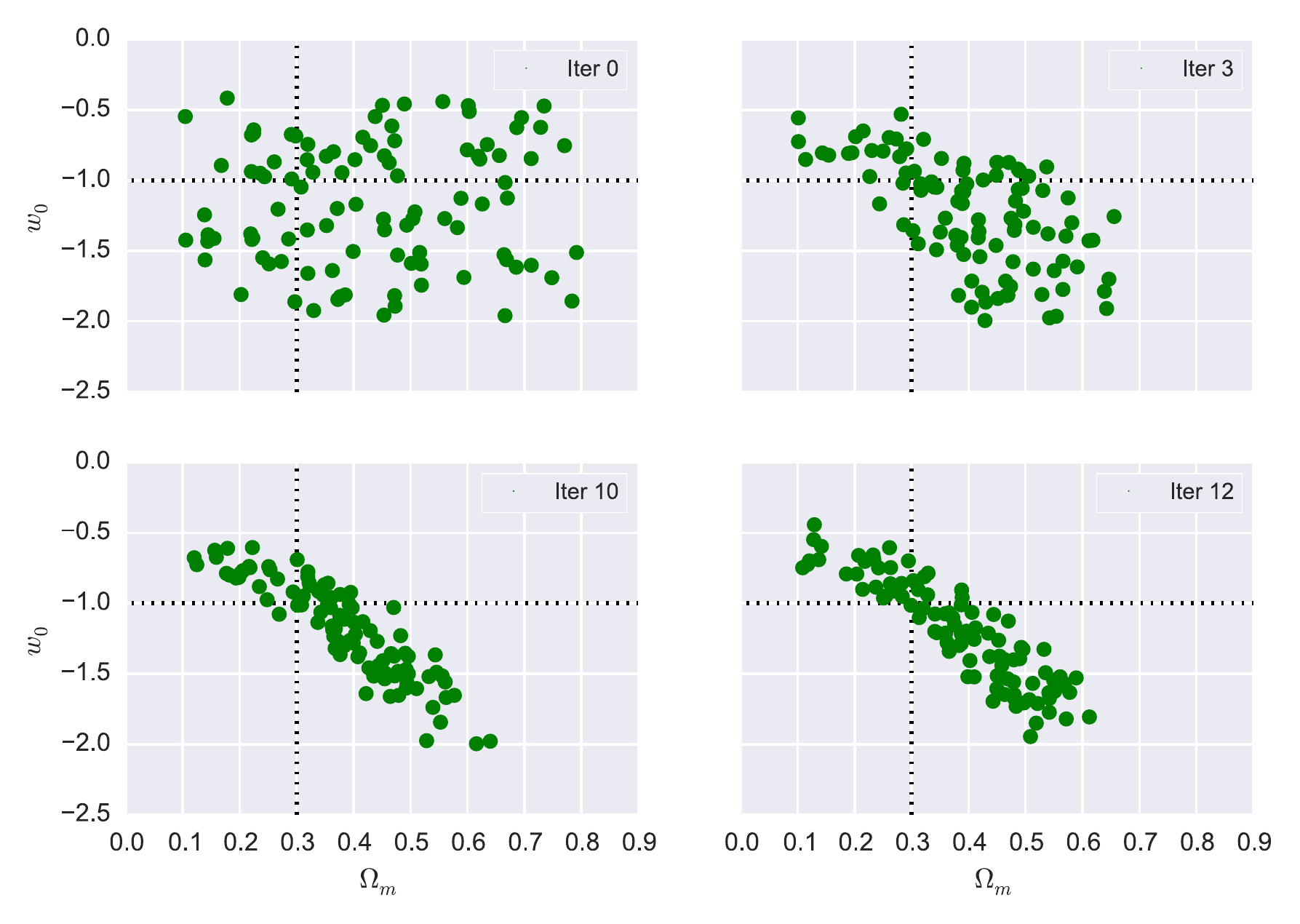}
\caption{The accepted parameter values for $\Omega_m$ and $w_0$ at four different iterations in the {\it superABC} sampling algorithm varying 5 parameters $(\Omega_m, w_0, \alpha, \beta, \delta M_0$) without including systematic uncertainties or Planck priors.
The ABC particles are represented by green circles at each iteration. The `true' parameters of the data are shown at the intersection of the dashed black lines.}
\label{fig:abc_particles}
\end{center}
\end{figure}

\subsubsection{Constraints using the Tripp metric with uniform priors}\label{sec:uniform}

\begin{figure}
\begin{center}
\includegraphics[height=3in,width=3.5in]{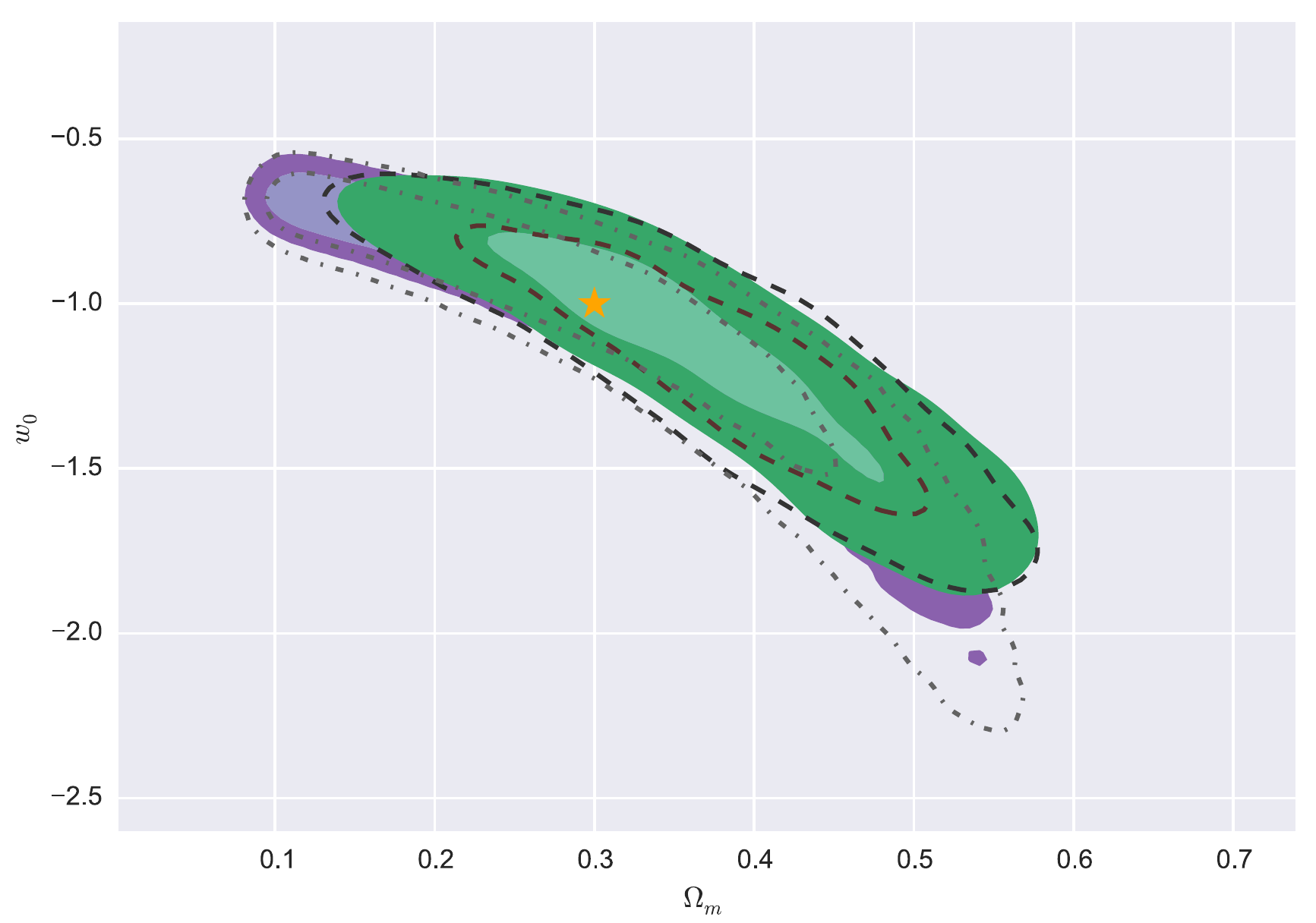}
\caption{The 1 and 2$\sigma$ contours for $\Omega_m$ and $w_0$ using the Tripp metric, discussed in Section \ref{sec:tripp}, {\it without} Planck priors.
The filled green contours show constraints obtained by varying 5 parameters $(\Omega_m, w_0, \alpha, \beta, \delta M_0)$ without including systematic uncertainties. The
dashed lines show  parameter constraints varying the same 5 parameters as well as four calibration uncertainty parameters.
The ``true" parameters of the data are represented by the yellow star.
The results from the MCMC sampler without (with)  systematic uncertainties are shown as filled purple contours (dot dashed lines).}
\label{fig:12sigma_NOplanck}
\end{center}
\end{figure}

In this section we present the parameter constraints using the Tripp metric with and without allowing for calibration uncertainties.
We vary five parameters, \{$\Omega_m, w_0, \alpha, \beta, \delta M_0$\}  in the {\it superABC } sampler with uniform priors.
In Fig. \ref{fig:abc_particles} we show the accepted parameters for $\Omega_m$ and $w_0$ at four different iterations in the sampler using 100 particles. Each particle is represented as a green circle in this figure and the `true' parameters of the data are shown as a dashed black line.
At iteration 0, the threshold for the distance metric is large and the accepted parameters are widely dispersed in parameter space. Note a threshold of infinity here would return a sample from the prior distribution for each of the parameters. As the iteration number increases, and the tolerance threshold decreases, the particles converge towards the true values of the parameters and occupy an extended ellipse.

In Fig. \ref{fig:12sigma_NOplanck} we show the 1 and 2$\sigma$ contours for $\Omega_m$ and $w_0$. The filled green contours show constraints obtained by varying five parameters $(\Omega_m, w_0, \alpha, \beta, \delta M_0)$ with statistical uncertainties only. The
dashed lines show  parameter constraints varying  nine parameters  (four of these are calibration uncertainty parameters).
The true parameters of the data are represented by the yellow star. In both cases we recover the true parameter value in the data within the 
1$\sigma$ error. Our marginalized 
constraints are $\Delta(\Omega_m) = \Omega_m^{\rm true}-\Omega_m^{\rm best \, fit} = -0.06\pm0.12, \Delta(w_0) =0.18\pm0.33 $ with statistical uncertainties and
$\Delta(\Omega_m) = -0.03\pm0.13, \Delta(w_0)= 0.17\pm0.37$ with systematics. The effect of including these four systematic uncertainties is to increase the uncertainty on
$w_0$ by $\sim$14\% and shift the best fit value closer to the true value.
We obtain constraints on $\Delta(\delta M_0)$ of $-0.009 \pm 0.005$ with statistical uncertainties and $-0.010\pm 0.008$ with both statistical and systematic uncertainties.

In Table \ref{table:1} we show the best fit values and 1$\sigma$ constraints on the SN standardization parameters $\alpha$ and $\beta$. We obtain approximately a 24\% error on $\alpha$ and a 7\% error on $\beta$ without systematic uncertainties which is relatively unchanged by the inclusion of systematic uncertainties. As the calibration uncertainties are largely degenerate with the cosmological parameter $w_0$, it is expected that most of the impact of including these uncertainties would be seen for this parameter.
Note the increased uncertainty on $\Omega_m$ with systematics compared to without is due to varying nine parameters with systematics as opposed to five parameters without.

We can compare the results from the {\it superABC sampler} with the results from using a MCMC sampler (Section \ref{sec:usual}) when systematic uncertainties are not taken into account. The MCMC results are 
$\Delta(\Omega_m) = -0.06 \pm  0.12,  \Delta(w_0)   =  0.05 \pm  0.34, \Delta(\alpha) = 0.007 \pm  0.020$
and  $\Delta(\beta) =   -0.40 \pm 0.19$. 
We achieve similar constraints on $\Omega_m$ and $w_0$ using the Tripp metric in this case, however our uncertainties
on $\alpha$ and $\beta$ are slightly larger than those obtained using MCMC (see Table \ref{table:1}) but in contrast to the MCMC results we achieve unbiased results for both parameters.
One interesting point to note is that we find non-zero covariance between the four parameters in both the results from the {\it superABC} sampler and MCMC. In particular using {\it superABC}  the correlation between $\alpha$ and $w_0$ is of the same order of magnitude as the correlation between $\alpha$ and $\beta$. This would suggest some degeneracy between the parameters in the likelihood model in the case of MCMC, and the metric in the case of {\it superABC}. This correlation is certainly not physical given the independence of cosmological parameters and SN parameters but may be a result of selection effects.\footnote{Alex Conley, private communication.} A detailed study of this effect is beyond the scope of this work.

In Appendix \ref{app:C}, when we fix $\Omega_m$, $w_0$ and $\delta M_0$ and only fit for 
$\alpha$ and  $\beta$ with the Tripp metric in {\it superABC}  we obtain significantly tighter constraints on these two parameters compared to varying five parameters simultaneously. 
As the covariance between the four parameters, $\Omega_m$, $w_0$, $\alpha$ and  $\beta$  is non-zero we would expect that allowing five parameters to  vary  increases the parameter uncertainties as found with the {\it superABC} sampler.

As discussed in Section \ref{sec:usual},  when including calibration uncertainties, 
the  MCMC method is accounting for systematics approximately while the ABC sampler is correctly including the effect of these at every point in parameter space.
As a result the  {\it superABC} sampler is fitting for nine parameters simultaneously 
while the  MCMC sampler is fitting for five parameters.  
To incorporate calibration uncertainties in the MCMC covariance matrix we follow the method in \citet{2011ApJS..192....1C}. Firstly SALT2 fit parameters are obtained for each SN  in the data set assuming no calibration systematics, then the data is refit assuming a maximum uncertainty of 0.01 mag. The difference in each of these fit parameters with and without uncertainties is calculated. Off diagonal elements in the covariance matrix are constants corresponding to a linear combination of the differences following equations 5 and 6  in \citet{2011ApJS..192....1C}.  In the {\it superABC} sampler calibration uncertainties in the
four bands are parameters with Gaussian priors, $\mathcal{N}(0,0.01)$, and so the impact of these systematics in the sampling method are allowed to vary from point to point in parameter space.
In Fig. \ref{fig:12sigma_NOplanck} we plot the results from the MCMC sampler with and without including systematic uncertainties as dot-dashed lines and purple filled contours respectively. 
The MCMC constraints accounting for systematics are 
$\Delta(\Omega_m) = -0.02\pm 0.12, \Delta(w_0) = 0.12 \pm 0.38, \Delta(\alpha) = 0.007 \pm 0.020$ and $\Delta(\beta) = -0.399 \pm  0.197$. 
Overall we see a $\sim 11$\%
 increase in the uncertainty on $w_0$ with systematics compared to without from the MCMC sampler which is smaller than the increase measured from the ABC sampler (14\%).
 We find that approximately including the effect of systematics using a fixed uncertainty in covariance matrix causes the MCMC sampling method to underestimate the uncertainties on $w_0$. 
 
 \begin{figure}
\begin{center}
\includegraphics[height=3in,width=3.5in]{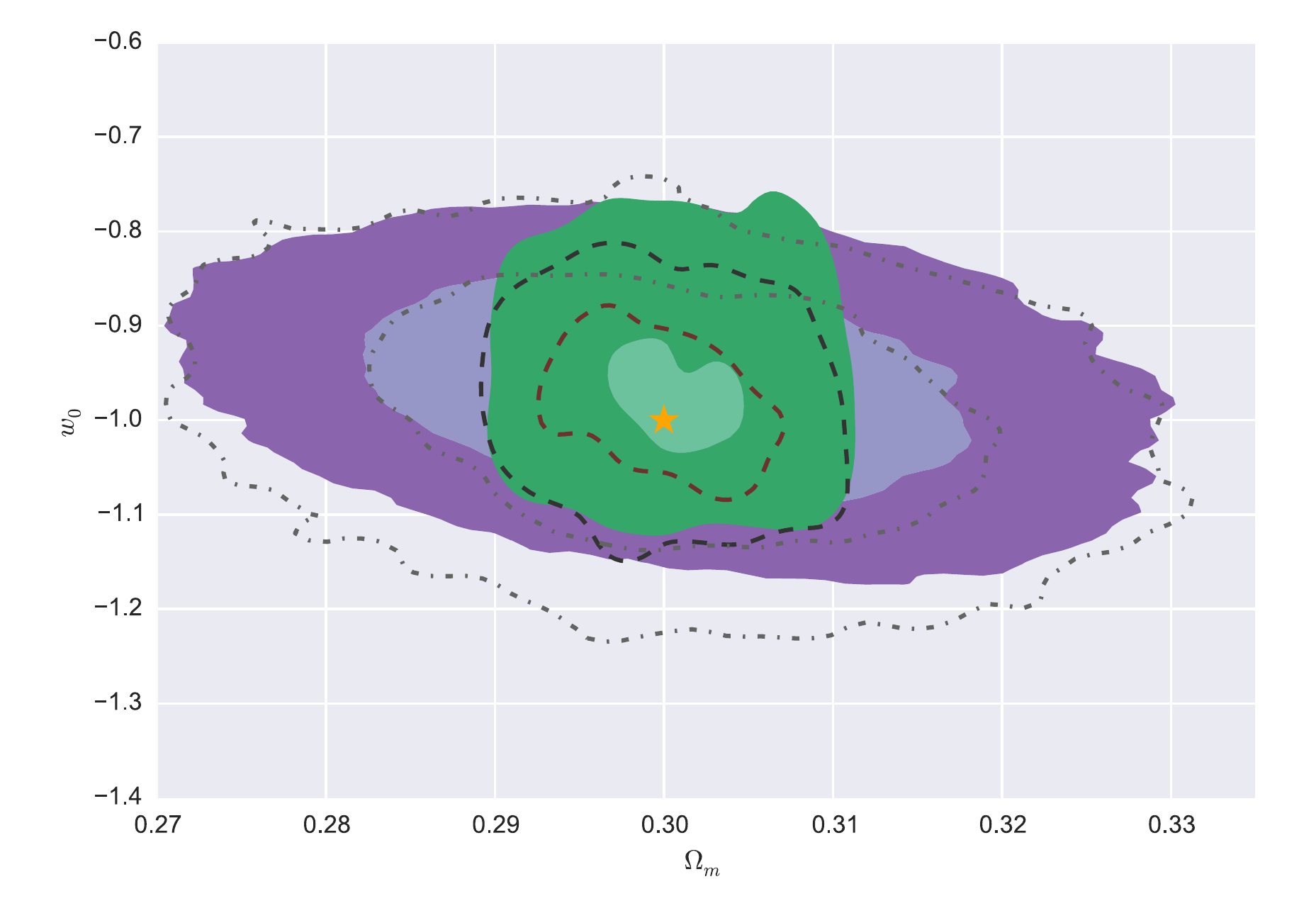} 
\caption{The 1 and 2$\sigma$ contours for $\Omega_m$ and $w_0$ using the Tripp metric {\it including} Planck priors on $\Omega_m$ and $w_0$.
The filled green contours show constraints obtained by varying  $(\Omega_m, w_0, \alpha, \beta, \delta M_0)$ without including systematic uncertainties. The
dashed lines show  parameter constraints varying the same 5 parameters as well as four calibration uncertainty parameters.
The ``true" parameters of the data are represented by the yellow star.
The results from the MCMC sampler without (with)  systematic uncertainties are shown as filled purple contours (dot dashed lines).}
\label{fig:12sigma_planck}
\end{center}
\end{figure}

 Note that we find a  bias in the best fit value of $\beta$ recovered using the MCMC sampler. This is possibly the result of neglecting selection effects in the likelihood model. As the selection function is consistently accounted for in the ABC simulations, and we are assuming we know the selection function that was applied to the data, there is no bias present in the {\it superABC} results.
 
\vspace{1cm}
\subsubsection{Constraints using the Tripp  metric with CMB priors}\label{sec:results_tripp2}

In this section we present the parameter constraints using the Tripp metric with and without allowing for calibration uncertainties and including  a Planck prior on $\Omega_m$ and $w_0$ as described in Section \ref{sec:priors}.
We vary five (nine) parameters, \{$\Omega_m, w_0, \alpha, \beta, \delta M_0$\}  in the {\it superABC } sampler when systematic uncertainties are neglected (included).

In Fig.\ \ref{fig:12sigma_planck} we show the 1 and 2$\sigma$ contours for $\Omega_m$ and $w_0$. The filled green contours show constraints obtained without accounting for systematic uncertainties. The
dashed lines show  parameter constraints varying  the calibration uncertainty parameters.
The true parameters of the data are represented by the yellow star. In both cases we recover the true parameter value in the data within the 
1$\sigma$ error. Our marginalized constraints are
 $\Delta(\Omega_m) = 0.001\pm0.006, \Delta(w_0) =-0.036\pm0.109 $ without systematics and
$\Delta(\Omega_m) = 0.0004\pm0.0062, \Delta(w_0)= -0.06\pm0.132$ with systematics. The effect of including these four systematic uncertainties is to increase the uncertainty on
$w_0$ by $\sim$17\%. 
We obtain constraints on $\Delta(\delta M_0)$ of $0.002 \pm 0.011$ without systematics and $0.013\pm 0.014$ with systematics.

The results from the {\it superABC} sampler can be compared with the results from using a MCMC sampler (Section \ref{sec:usual}) when systematic uncertainties are not taken into account. 
In Fig.\ \ref{fig:12sigma_planck} we plot the results from the MCMC sampler with and without including systematic uncertainties as dot-dashed lines and purple filled contours respectively.
The MCMC results are $\Delta(\Omega_m) = 0.000 \pm 0.012,  \Delta(w_0)   =      -0.03 \pm 0.082, \Delta(\alpha) =0.008 \pm  0.020$
and  $\Delta(\beta) =   -0.41 \pm  0.19$. 
From this plot is it clear that the ABC 1 and 2$\sigma$ contours are a less symmetric and smaller than those from the MCMC sampler.
This is possibly due to the lack of any Gaussian likelihood assumption in the case of ABC and the fact that the prior dependence in each algorithm is very different.
We achieve tighter constraints on $\Omega_m$ and slightly larger constraints on $w_0$ using the Tripp metric in this case. 
When including systematic uncertainties  the  {\it superABC} sampler is correctly accounting for these uncertainties and fits for nine parameters simultaneously,
while the  MCMC sampler uses an estimated uncertainty in the covariance matrix  and fits for five parameters.  The MCMC constraints accounting for systematics are 
$\Delta(\Omega_m) = -0.0005 \pm  0.0119, \Delta(w_0) = -0.006 \pm  0.097, \Delta(\alpha) = 0.0077 \pm  0.0201$ and $\Delta(\beta) = -0.40 \pm 0.20$. 
Overall we see a $\sim 15$\%
 increase in the error on $w_0$ with systematics compared to without from the MCMC sampler which is smaller then the increase found using the ABC sampler (17\%). 
 As found in Section \ref{sec:uniform} incorporating systematic errors approximately as in the MCMC sampler causes a slight underestimation of the uncertainties on $w_0$.
 Including systematics uncertainties  slightly shifts the 1 and 2$\sigma$ contours towards more negative values of $w_0$ for both samplers.
We find similar uncertainties 
on $\alpha$ and $\beta$ with both the {\it superABC} and the MCMC sampler, however as before we find a bias in the best fit value of $\beta$ from MCMC which is not seen with the ABC sampler (see Table \ref{table:1}).  This is likely to be the result of neglecting selection effects in the likelihood model in the MCMC sampler.

\begin{table*}
\caption{The difference between the `true' and best fit values for the
parameters \{$\Omega_m, w_0, \alpha, \beta$\} obtained from the {\it superABC} sampler using 100 particles for 18 iterations.
E.g.\ In this table $\Delta(\Omega_m) = \Omega_m^{\rm true} - \Omega_m^{\rm best \, fit}$. 
The  uncertainties shown are the standard deviation of the 1D marginalized pdfs. The true values of the parameters in the data are $\Omega_m=0.3, w_0=-1.0, \alpha=0.14$ and $\beta = 3.2$. 
The number in brackets for the Tripp metric with Planck priors and without systematic uncertainties represents the standard deviation on the 1$\sigma$ error reported amongst 10 different realizations of the data.
The last four rows show the results from an MCMC sampler used on the same data.
\label{table:1}}
\begin{center}
\begin{tabular}{ |p{1.3cm}||p{1.1cm}|p{3.9cm}|p{2.5cm}|  p{2.5cm}|  p{2.5cm}|  p{2.5cm}| }
 \hline
\hline
 & sampler & priors on \{$\Omega_m, w_0$\} & $\Delta(\Omega_m)$  & $\Delta(w_0)$ & $\Delta(\alpha)$ & $\Delta(\beta)$\\
\hline 
\multirow{3}{4em}{stat only} & Tripp & $\mathcal{U}(0.05,0.95), \,\,    \mathcal{U}(-2.5,-0.2) $ & $-0.06\pm0.12$ &$0.18\pm0.33$ &$0.010\pm0.031$ & $-0.13\pm0.25$ \\ 
& Tripp  & Planck &  $0.001\pm0.006(0.0001)$ &$-0.036 \pm0.109 (0.006)$ &$-0.009\pm0.028 (0.0007)$&$-0.005\pm0.280 (0.011)$  \\ 
& Light Curve &  $\mathcal{U}(0.05,0.95), \,\,   \mathcal{U}(-2.5,-0.2) $ & $-0.09\pm0.10$ & $0.32\pm0.33$ & $0.003\pm0.021$ & $0.05\pm0.29$  \\ 
& Light Curve & Planck  & $0.002\pm0.012$& $-0.05\pm0.06$ & $0.009\pm0.018$  & $0.07\pm0.22$ \\ 
\hline
\multirow{3}{4em}{with sys} & Tripp & $\mathcal{U}(0.05,0.95), \,\,  \mathcal{U}(-2.5,-0.2) $ & $-0.03\pm0.13$ & $0.17\pm0.37$ &$0.001\pm0.040$ &$-0.17\pm0.38$   \\ 
& Tripp  & Planck &$0.0004\pm0.0062$  &$-0.06\pm0.132$ &$-0.01\pm0.02$ &$-0.08\pm0.28$ \\ 
\hline
\hline
\hline
\multirow{3}{4em}{stat only} & MCMC & $\mathcal{U}(0.05,0.95), \,\,  \mathcal{U}(-2.5,-0.2) $ & $-0.06\pm0.12$ & $0.05\pm0.34$ &$0.007\pm0.020$ &$-0.40\pm0.19$   \\ 
& MCMC  & Planck &$0.000\pm0.012$  &$-0.03\pm0.082$ &$0.008\pm0.020$ &$-0.41\pm0.19$ \\ 
\hline
\multirow{3}{4em}{with sys} & MCMC & $\mathcal{U}(0.05,0.95), \,\,  \mathcal{U}(-2.5,-0.2) $ & $-0.02\pm0.12$ & $0.12\pm0.38$ &$0.0076\pm0.0204$ &$-0.399\pm0.197$   \\ 
& MCMC  & Planck &$-0.0005\pm0.0119$  &$-0.006\pm0.097$ &$0.0077\pm0.0201$ &$-0.40\pm0.20$ \\ 
\hline
 \hline
\end{tabular}
\end{center}
\end{table*}

\begin{figure}
\begin{center}
\includegraphics[height=3in,width=3.5in]{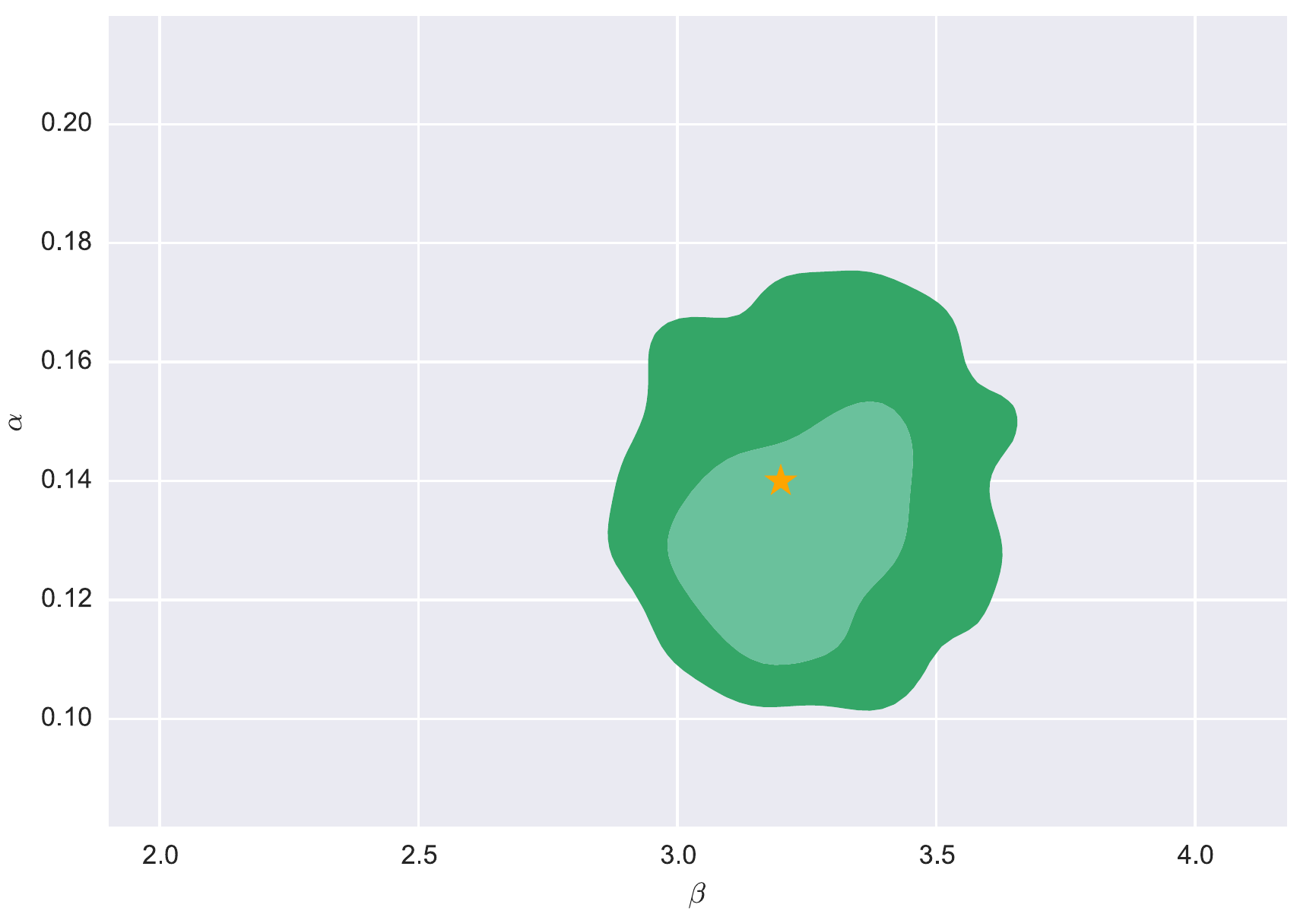}
\caption{The 1 and 2$\sigma$ contours for $\alpha$ and $\beta$ using the Tripp metric with Planck priors on
$\Omega_m$ and $w_0$. 
The filled green contours show constraints obtained by varying $(\Omega_m, w_0, \alpha, \beta, \delta M_0)$ as well as four calibration uncertainty parameters.
The ``true" parameters of the data are represented by the yellow star.}
\label{fig:alpha_beta}
\end{center}
\end{figure}

In Fig.\ \ref{fig:alpha_beta} we plot the 1 and 2$\sigma$ contours for the SN standardization parameters $\alpha$ and $\beta$ allowing for systematic uncertainties. We recover the correct value of $\alpha$ and $\beta$ within the 1$\sigma$ error bar. As these constraints were obtained by fitting for 9 parameters simultaneously in the {\it superABC} sampler it is not expected that we should achieve the same precision on the standardization parameters which is possible with SALT2mu maximum likelihood technique which fits for 2 parameters.
 In Appendix \ref{app:C} we show that the Tripp metric can constrain the standardization parameters with comparable precision to SALT2mu when the cosmological parameters are fixed.

\subsection{Constraints using the Light Curve metric}\label{sec:results_lc}

\begin{figure}
\begin{center}
\includegraphics[height=3in,width=3.5in]{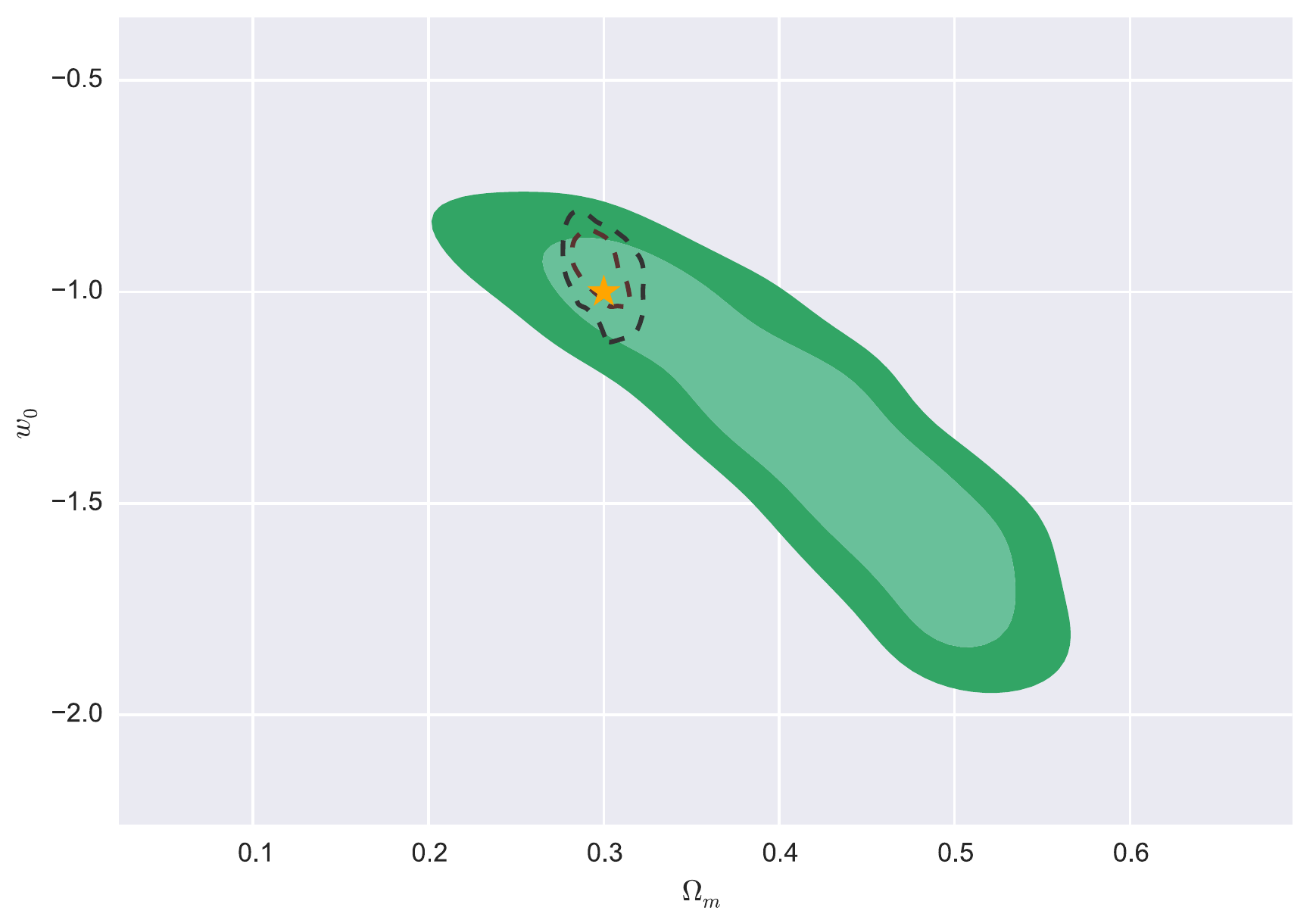}
\caption{The 1 and 2$\sigma$ contours for $\Omega_m$ and $w_0$ using the Light Curve metric, discussed in Section \ref{sec:lc_metric}, with Planck and uniform priors
are shown as the dashed and solid contours respectively.
Both sets of constraints shown were obtained by varying $(\Omega_m, w_0, \alpha, \beta, \delta M_0)$ without including systematic uncertainties.
The ``true" parameters of the data are represent by the yellow star.}
\label{fig:lc_results}
\end{center}
\end{figure}

In Fig.\ \ref{fig:lc_results}  the 1 and 2$\sigma$ contours for $\Omega_m$ and $w_0$ using the Light Curve metric, discussed in Section \ref{sec:lc_metric}, with Planck and uniform priors
are shown as the dashed and solid contours respectively. These constraints were obtained varying 
$(\Omega_m, w_0, \alpha, \beta, \delta M_0)$ without including systematic uncertainties. 
Without systematics we obtain an $\sim7$\% error on $w_0$ using the Light Curve metric compared to $11$\% using the Tripp metric.

As described in Section \ref{sec:lc_metric} the Light Curve metric does not depend on the Tripp relation where the parameters 
$\alpha$ and $\beta$ are defined. 
For both the constraints with Planck and uniform priors we recover the true parameters of the data (yellow star) within the 1$\sigma$ error. The full results are given in Table \ref{table:1}.
We obtain constraints on $\Delta(\delta M_0)$ of $-0.009 \pm 0.01$ with uniform priors and $-0.013\pm 0.011$ with Planck priors.
Overall the effect of including Planck priors decreases the uncertainties on $\Omega_m$ and $w_0$ by
$\sim$87\% and  $\sim$79\%  respectively without systematics.

\vspace{0.5cm}
\section{Summary of results}\label{sec:summary}

In the previous section we  presented results from the {\it superABC} sampler using both the Tripp and the Light Curve distance metrics; and where appropriate, we have provided results obtained by using the JLA likelihood method in an MCMC sampler. 
Note that throughout this paper we have used the term 'MCMC results'  to refer to a choice of likelihood, model and method of including systematics (which may be estimated or exact) in a MCMC analysis.

Overall there were several motivations for considering both the Tripp metric in addition to the Light Curve metric. Firstly as computational speed can be an issue with any ABC method, we found that the Tripp metric is approximately 30\% faster  than the Light Curve metric in estimating the posterior distribution for the same number of walkers in the same number of iterations.
Secondly, in order to compare with earlier methods which use the Tripp relation in the likelihood, it is appropriate to construct an ABC distance metric which also uses this relation. Finally, in constraining the parameters $\alpha$ and $\beta$, which are defined in the Tripp relation, we are able to compare the constraints obtained using a metric which is based on this relation and a metric which only uses the light curve fluxes directly.

The key results of this paper are the use of two distinct and novel distance metrics, in a sampler that uses a forward model simulation for every proposed point in parameter space, which can consistently incorporate the systematic effect of calibration uncertainties.
Previous MCMC approaches either neglect  systematics or  include
their effects  using approximations of a fixed uncertainty which is added to the covariance matrix prior to sampling.
In this paper we present  MCMC results which include
 calibration uncertainties using the approximation given in \citet{2011ApJS..192....1C}, which is part of the JLA likelihood. In this approximation calibration uncertainties are included in the off
 diagonal elements in the covariance matrix as a fixed constant offset of
 0.01 mag in each of the four bands.
 This likelihood approach does not include a forward model where the effects of systematics can be parametrized and marginalized over as in the {\it superABC} sampler.
As we discuss in Section \ref{sec:usual} when correctly accounting  for systematics in the  {\it superABC} sampler we are fitting for nine parameters simultaneously, while in
the  MCMC sampler we are fitting for five parameters. 
 In summary, current MCMC methods rely on estimated systematics and covariances and may  be
 over or underestimating 
 the impact of these systematics on cosmological parameter constraints. ABC methods allow us to correctly include systematics at every point in parameter space during sampling.
 Contrasting our ABC results with MCMC allows us to determine whether or not the effect of systematic uncertainties has been under or overestimated in the MCMC approach. It also allows us to test for any biases in the best fit values obtained which may  result from neglecting selection effects.

From Table \ref{table:1} we can compare the constraints obtained using both the Tripp and the Light Curve metric without accounting for systematic uncertainties. Overall we find consistent  1$\sigma$ constraints on $\Omega_m$ and $w_0$  using either metric with a slightly larger error for $w_0$ using the Tripp metric with Planck priors ($\sim 11$\%) compared to using the Light Curve metric with the same priors ($\sim7$\%). 
We can compare the results from the {\it superABC} sampler with the results from using  the JLA likelihood in a MCMC sampler in the case of uniform priors when systematic uncertainties are not taken into account.
  From the `stat only' rows in Table \ref{table:1} it is clear that the Tripp metric obtains similar uncertainties on $\Omega_m$ to the MCMC sampler with and without CMB priors. With the CMB priors the ABC sampler obtains slightly tighter constraints on $\Omega_m$ and weaker constraints on $w_0$.
 The agreement between these results and our results shown in Table \ref{table:1} 
shows that both {\it superABC} metrics are able to recover the `true' cosmological parameters with similar precision
to the MCMC sampler but without biased best fit values. We find a bias in the best fit value of $\beta$ from MCMC which is not seen with the ABC sampler. This is possibly the result of neglecting selection effects in the likelihood model in MCMC \citep[see][for a likelihood based approach to this using forward model simulations]{bambis}.

The constraints from {\it superABC} on the standardization parameters $\alpha$ and $\beta$, 
in Table \ref{table:1},  are similar using either the Tripp or Light Curve metric with uniform priors. However in Appendix \ref{app:C} we show that the Tripp metric can produce tighter constraints on $\alpha$ and $\beta$ once the other parameters are fixed (in analogy to the SALT2mu maximum likelihood method). 
There is no explicit reference to $\alpha$ and $\beta$ in the Light Curve metric but the distribution of differences, which we compute as part of the metric, is sensitive to these two parameters i.e. incorrect proposed values of $\alpha$ and $\beta$ will produce simulated light curves which look very different to the data light curves.
However, in our tests the Light Curve metric is unable to match the constraints on $\alpha$ and $\beta$ from the Tripp metric . Even when all other parameters are fixed
the Light Curve metric consistently achieves uncertainties on $\alpha$ and $\beta$ which are $\sim 70-80$\% higher than those from SALT2mu. This is not completely unexpected given that the Light Curve metric makes no explicit reference to the Tripp relation where $\alpha$ and $\beta$ are defined in terms of the SALT-II fitted parameters.

As the Tripp metric is better suited to constraining the SN standardization parameters  we choose to use only this  metric when evaluating the effect of systematic uncertainties.
Using the Tripp metric we find that accounting for calibration uncertainties will increase the uncertainty on $w_0$, by $\sim$17\% with Planck priors and
$\sim$14\% with uniform priors, compared to not including these systematics. 
As mentioned previously if we want to include these in an MCMC method, we can estimate their impact using e.g.\ an MCMC sampler with a JLA style likelihood (equation 15 in \citet{2014A&A...568A..22B}). 
In Table \ref{table:1} we see that the MCMC sampler results show an 11\% increase in the uncertainties  on $w_0$ with uniform priors and a 15\% increase with a Planck priors compared to not including these systematics. Both of these MCMC estimates  are  smaller then the corresponding predictions from the ABC sampler (14\% for uniform priors and 17\% for Planck priors) and would suggest that the MCMC method has slightly underestimated the impact of calibration uncertainties for this data set.

The main assumption in trusting the results of the {\it superABC} sampler is that the forward model simulation accurately includes the effects of systematics, and consistently accounts for their impact on the light curves and fluxes, as we vary the cosmological parameters. If this is not the case then in any ABC analysis we will not recover the correct posterior distribution. As we allow the parameters for calibration uncertainty to vary along with the cosmological parameters during sampling we are naturally accounting for correlations and degeneracies between parameters. In this paper,  systematic uncertainties are included as a fixed value in the covariance matrix in the likelihood and are not allowed to vary. This assumption is only correct if (1) we are certain that the systematics are not correlated with the cosmological parameters and (2) we are certain about the size of these uncertainties. We know that calibration uncertainties are degenerate with cosmological parameters so assuming (1) is not correct here and using a fixed 0.01 mag offset is an estimate so we are not certain about (2).

It is important to note that our results including Planck priors cannot be strictly compared to a MCMC analysis where a joint Planck and SN likelihood are sampled from simultaneously, e.g.\ the JLA results in \citet{2014A&A...568A..22B} used CMB data and SN data in two different likelihoods. 
The reason for this is that we do not use likelihoods in the ABC sampler and any joint probe analysis  in ABC would need forward model simulations for all datasets considered. In this paper the {\it superABC} sampler  uses Planck priors but we are not using the Planck likelihood or the CMB data during sampling. 
Note we do not mean to imply here that the ABC method is insensitive to priors,  only that using a conservative CMB prior and sampling with SN only data is not the same as jointly sampling with CMB and SN data.

As a final point to summarize our results we find that there are two clear cases where using one ABC metric would be more optimal than the other.
If the primary interest is speed of computation and constraining the standardization parameters $\alpha$ and $\beta$ then we would recommend using the Tripp metric where these parameters
are explicitly defined and it is slightly faster to evaluate. If someone wants to avoid using the SALT2 fit parameters and to use the light curve fluxes directly we would recommend using the Light Curve metric.
Once the forward model simulation is extended to allow other systematic parameters to vary then  both of these metrics would need to be tested to see if they are sensitive to the new parameters we wish to constrain, i.e. we would need to test that the summary statistic used in the metric is sufficient for these new parameters (Section \ref{sec:metric}).

\section{Discussion}\label{sec:discussion}

In this paper we have presented a new analysis package, {\it superABC}, for obtaining cosmological constraints from SNe using Approximate Bayesian Computation.
The {\it superABC} sampler is applied to a simulated data set of $\sim$1000 SNe based on the  first DES SNe program season.
A key feature of the analysis method presented here is the use of two distinct metrics, the `Tripp' and `Light Curve' metric, which allow us to compare the forward model simulated data to the observed data set without likelihood assumptions.
The Tripp metric is based on the Tripp relation \citep{1998A&A...331..815T} 
and uses the SALT-II framework fitted light-curve parameters for each SN.  The Light Curve metric compares the data and simulated  light curve fluxes output from {\sc SNANA} for each observation, for each event, at every proposed point in parameter space.

The method presented in this paper represents a completely new approach to constraining cosmological parameters using SN data without any  likelihood assumptions in a  framework which can naturally incorporate systematic uncertainties. In this initial methods paper we focus on the effects of 
calibration uncertainties and priors on cosmological constraints. An obvious next step is to extend this analysis and e.g.\ fit for the parameters that describe the intrinsic populations of SN Ia color and stretch that correlate with luminosity and  parameters that account for  contamination from core collapse SNe. This will require us to devise new ABC metrics that are sensitive to any variables used to parametrize the systematics and is beyond the scope of this initial work \citep[however see][for a likelihood based analysis of these effects which uses forward model simulations]{bambis}.

The  ABC method presented here could account for core collapse contamination in the data sample using e.g.\ the light curve analysis
software `Photometric SN IDentification' (PSNID) \citep{2011ApJ...738..162S}  to give an estimated contamination probability. In the {\it superABC} sampler, at every point in parameter space we would generate SN  Ia and core collapse light curves
with either a fixed or floating contamination rate as given by PSNID. In this way our forward model simulation would naturally incorporate this contaminant and its effects on the cosmological parameters of interest would be marginalized over.
This ABC method could also be applied to a photometric analysis  where we have photometrically-classified SN but with spectroscopic redshifts of the host galaxy. \citet{2016arXiv160406138G} presented an automated algorithm which can be run on the catalog data and matches SNe to their hosts with 91\% accuracy.
With an estimation of the host-SN matching accuracy for the data set under consideration,  this could be incorporated into the forward model simulation in {\it superABC} by e.g.\ increasing the redshift  uncertainty on a given percentage of the SN at every point in parameter space. Again the percentage accuracy could be a fixed amount or could be treated as a free parameter with  a prior range based on the algorithm of \citet{2016arXiv160406138G}.
We 
shall address these issues in a  future study.

\section{Acknowledgements}
We thank Rick Kessler for support running SNANA and Tamara Davis, Eve Kovacs, Tesla Jeltema, Scott Dodelson, Josh Frieman, 
Lorne Whitewall, Gary Bernstein, Dan Scolnic, Adam Riess, Alex Conley, Marc Betoule and Chad Schaffer for useful discussions and comments.
EJ is supported by Fermi Research Alliance, LLC under the U.S. Department of Energy under contract No. DE-AC02-07CH11359. 
Operated by Fermi Research Alliance, LLC under Contract No. De-AC02-07CH11359 with the United States Department of Energy.
MS and RW was supported by DOE grant DE-FOA-0001358 and NSF grant AST-1517742.
This paper has gone through internal review by the DES collaboration.
Funding for the DES Projects has been provided by the U.S. Department of Energy, the U.S. National Science Foundation, the Ministry of Science and Education of Spain, 
the Science and Technology Facilities Council of the United Kingdom, the Higher Education Funding Council for England, the National Center for Supercomputing 
Applications at the University of Illinois at Urbana-Champaign, the Kavli Institute of Cosmological Physics at the University of Chicago, 
the Center for Cosmology and Astro-Particle Physics at the Ohio State University,
the Mitchell Institute for Fundamental Physics and Astronomy at Texas A\&M University, Financiadora de Estudos e Projetos, 
Funda{\c c}{\~a}o Carlos Chagas Filho de Amparo {\`a} Pesquisa do Estado do Rio de Janeiro, Conselho Nacional de Desenvolvimento Cient{\'i}fico e Tecnol{\'o}gico and 
the Minist{\'e}rio da Ci{\^e}ncia, Tecnologia e Inova{\c c}{\~a}o, the Deutsche Forschungsgemeinschaft and the Collaborating Institutions in the Dark Energy Survey. 
The Collaborating Institutions are Argonne National Laboratory, the University of California at Santa Cruz, the University of Cambridge, Centro de Investigaciones Energ{\'e}ticas, 
Medioambientales y Tecnol{\'o}gicas-Madrid, the University of Chicago, University College London, the DES-Brazil Consortium, the University of Edinburgh, 
the Eidgen{\"o}ssische Technische Hochschule (ETH) Z{\"u}rich, 
Fermi National Accelerator Laboratory, the University of Illinois at Urbana-Champaign, the Institut de Ci{\`e}ncies de l'Espai (IEEC/CSIC), 
the Institut de F{\'i}sica d'Altes Energies, Lawrence Berkeley National Laboratory, the Ludwig-Maximilians Universit{\"a}t M{\"u}nchen and the associated Excellence Cluster Universe, 
the University of Michigan, the National Optical Astronomy Observatory, the University of Nottingham, The Ohio State University, the University of Pennsylvania, the University of Portsmouth, 
SLAC National Accelerator Laboratory, Stanford University, the University of Sussex, Texas A\&M University, and the OzDES Membership Consortium.
The DES data management system is supported by the National Science Foundation under Grant Number AST-1138766.
The DES participants from Spanish institutions are partially supported by MINECO under grants AYA2012-39559, ESP2013-48274, FPA2013-47986, and Centro de Excelencia Severo Ochoa SEV-2012-0234.
Research leading to these results has received funding from the European Research Council under the European Union?s Seventh Framework Programme (FP7/2007-2013) including ERC grant agreements 
 240672, 291329, and 306478.
We are grateful for the support of
the University of Chicago Research Computing Center.

\appendix

\section{The kernel and weighted covariance}\label{app:A}

In  Algorithm \ref{alg_1}  we assign weights, $w_{i,t}$, to each particle $i$ at iteration $t$ as
\ba
w_{i,t} \leftarrow  \frac{\pi(\theta_{i,t})}{\sum^{N}_{j=1}w_{j,t-1}\mathcal{K}(\theta_{j,t-1}|\theta_{i,t}, \mathcal{C}_{t-1})} \,,
\ea
where the Gaussian kernel, $\mathcal{K}$,  for parameter set $\theta_{k,t-1}$ given the parameter set $\theta_{l,t}$, is 
\ba
\mathcal{K}(\theta_{k,t-1}|\theta_{l,t}, \mathcal{C}_{t-1}) &=& (2\pi)^{-1/2}(|\mathcal{C}_{t-1}|)^{-1/2}e^{ -\frac{1}{2}\left(\theta_{k,t-1} - \theta_{l,t}\right)^{{\bf T}} \mathcal{C}_{t-1}^{-1} \left( \theta_{k,t-1} - \theta_{l,t}\right)  } \, ,
\ea
where the covariance is the weighted covariance matrix amongst all the particles, 
\ba
\mathcal{C} &=&  \frac{\sum_{i=1}^N w_i}{(\sum_{i=1}^N w_i)^2 - \sum_{i=1}^N w_i^2} \sum_{i=1}^N w_i (\theta_i - \bar{\theta})^{{\bf T}} (\theta_i - \bar{\theta}) \, ,
\ea
and
\ba
\bar{\theta} &=& \frac{\sum_{i=1}^N w_i \theta_i}{\sum_{i=1}^N w_i}  \, .
\ea

\section{The Light Curve metric reference pdf} \label{app:B}

\begin{figure}[H]
\begin{center}
\includegraphics[height=3.5in,width=4.in]{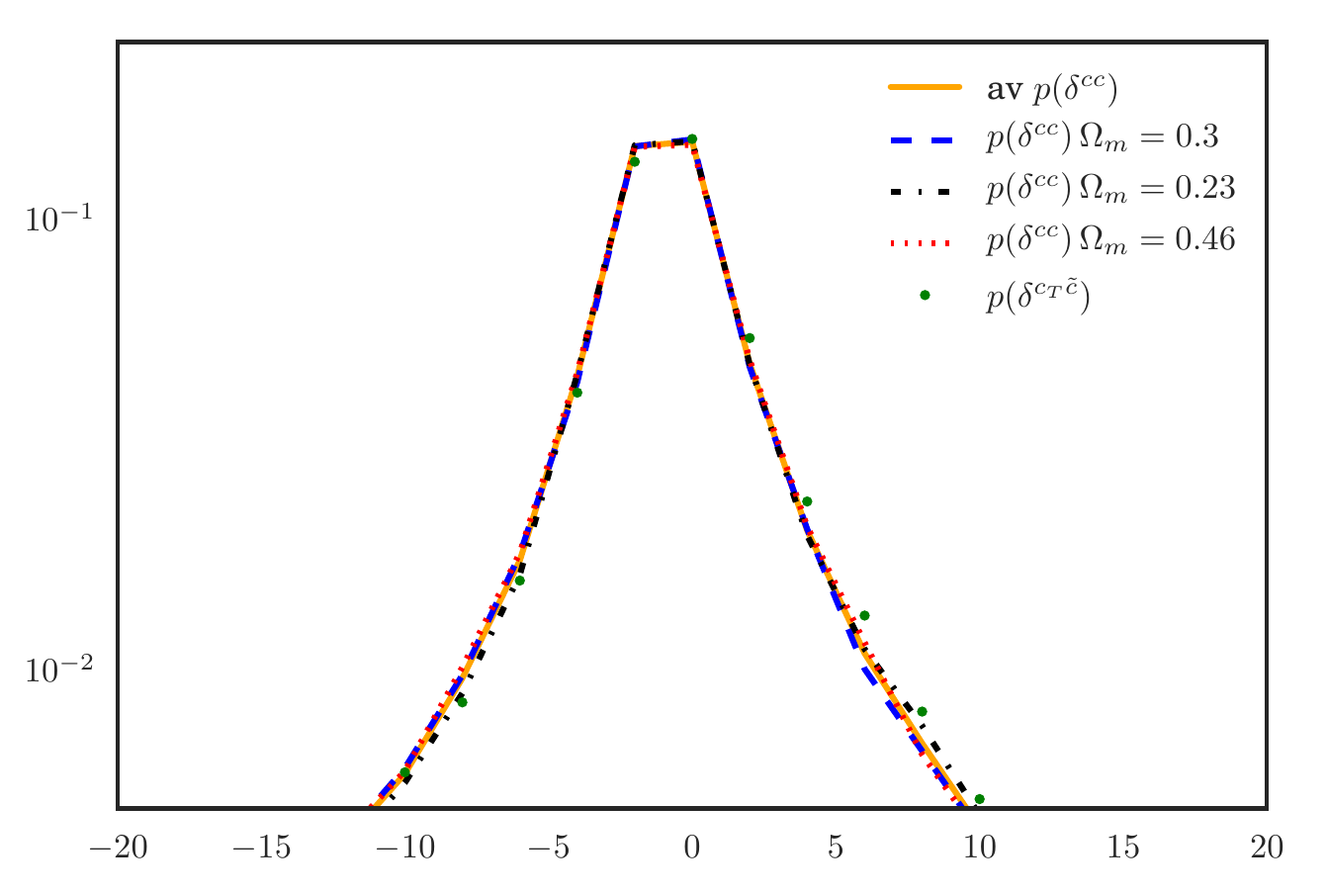}
\caption{The normalized reference difference distribution,  $p(\delta^{cc})$, using values of $\Omega_m = 0.3$ (blue dashed),
 $\Omega_m = 0.23$ (black dot dashed), $\Omega_m = 0.46$ (red dotted) and the average of these three (orange).
 The pdf $p(\delta^{c_T\tilde{c}})$, where the mock data and simulation had {\it different} cosmologies ($\Omega_m=0.3$ and $\Omega_m = 0.32$ respectively),
  is shown as green dots. Note the distribution appears flat at the peak due to binning effects.}
\label{fig:ref_pdf}
\end{center}
\end{figure}

In Section \ref{subsec:ref_sim_data_pdf} the method for creating a reference difference pdf is presented. In practice our reference pdf is an average over 3 pdf's each in a different cosmology (different $\Omega_m$). In Fig.\ \ref{fig:ref_pdf} we show the normalized reference difference distribution,  $p(\delta^{cc})$, using values of $\Omega_m = 0.3$ (blue dashed),
 $\Omega_m = 0.23$ (black dot dashed), $\Omega_m = 0.46$ (red dotted) and the average of these three (orange). Note all of these distributions appear flat at the peak due to binning effects. Although there are small differences between these three distributions, these difference are not as large as the difference between the average pdf (orange)
 and the pdf $p(\delta^{c_T\tilde{c}})$, where the mock data and simulation had {\it different} cosmologies ($\Omega_m=0.3$ and $\Omega_m = 0.32$ respectively).

\begin{figure}[H]
\begin{center}
\includegraphics[height=3in,width=3.5in]{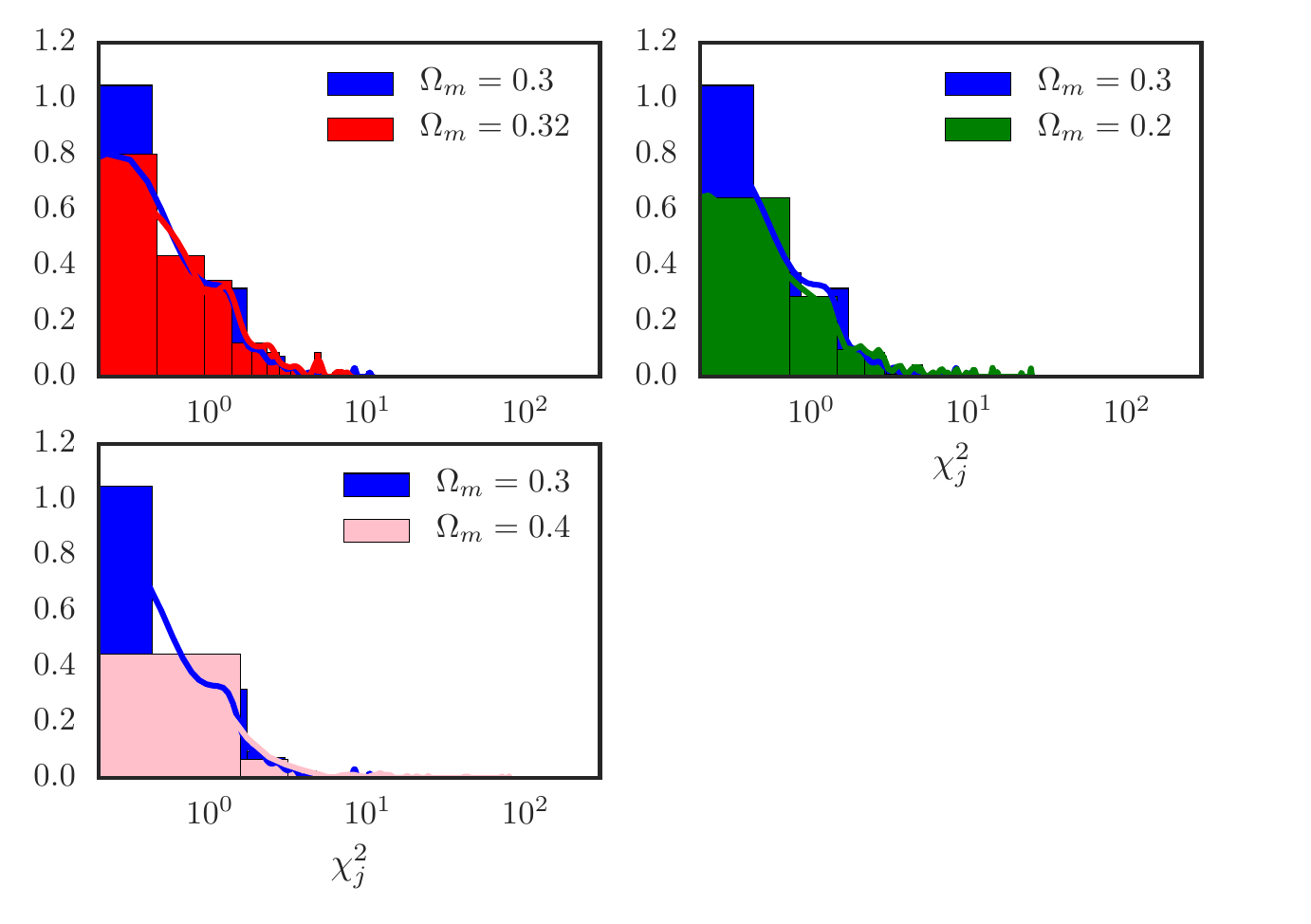}
\caption{The normalized distributions of  $\chi^2_j$ (Eq. \ref{eq:lc_metric2}) where the data has a fixed cosmology $\Omega_m = 0.3$ and in each case the simulated
events are generated in trial cosmologies of
$\Omega_m^* = 0.2,0.3,0.32,0.4$}.
\label{fig:teststatistic}
\end{center}
\end{figure}

As noted in Section \ref{sec:lc_metric_sampling} the Light Curve metric is distributed according to the $\chi^2$ distribution with $N_{\rm bins}$ degrees of freedom.
 So we can also state a p-value\footnote{For anyone not familiar with p-values, here it represents the probability of observing a test statistic at least as extreme in a chi-squared distribution with the given number of degrees of freedom. } for the $\chi^2$value we obtain.
In Fig.\ \ref{fig:teststatistic} we show the normalized distributions of $\chi^2_j$ (Eq. \ref{eq:lc_metric2}) where the simulated  light curve events are generated in trial cosmologies of
$\Omega_m^* = 0.2, 0.3,0.32,0.4$.  In this figure the curves represent a smoothed version of the histograms shown.
The  `true' parameter value in the data is $\Omega_m = 0.3$.
Note the same reference difference pdf was used in each case. There is a clear distinction between these four distributions. The distribution resulting from 
 a  simulation with proposed parameter value $\Omega_m^*=0.3$ is more peaked around zero and has a shorter tail than the distribution using a 
simulation with proposed parameter value $\Omega_m^*=0.4$. Increasing $\Omega_m^*$, while holding all other cosmological parameters fixed, causes a clear trend in spreading out the distribution and extending the tails. 
In this figure summing over the blue distribution  would represent the ABC metric, $\sum_j \chi^2_j$ (Eq. \ref{eq:lc_metric}), when the data and simulated events have the same cosmology, i.e.\ the ABC sampler has proposed a point in parameter space which exactly matches the data.
For the other distributions plotted in the three panels the distance metric  $\sum_j \chi^2_j$ would be larger than the metric for the blue distribution.
In this simple example the value of  $\sum_j \chi^2_j$ for the blue distribution ($\Omega_m^*=0.3$) represents the minimum threshold that can be achieved by the ABC sampler.

\begin{figure}[H]
\begin{center}
\includegraphics[height=2in,width=2.5in]{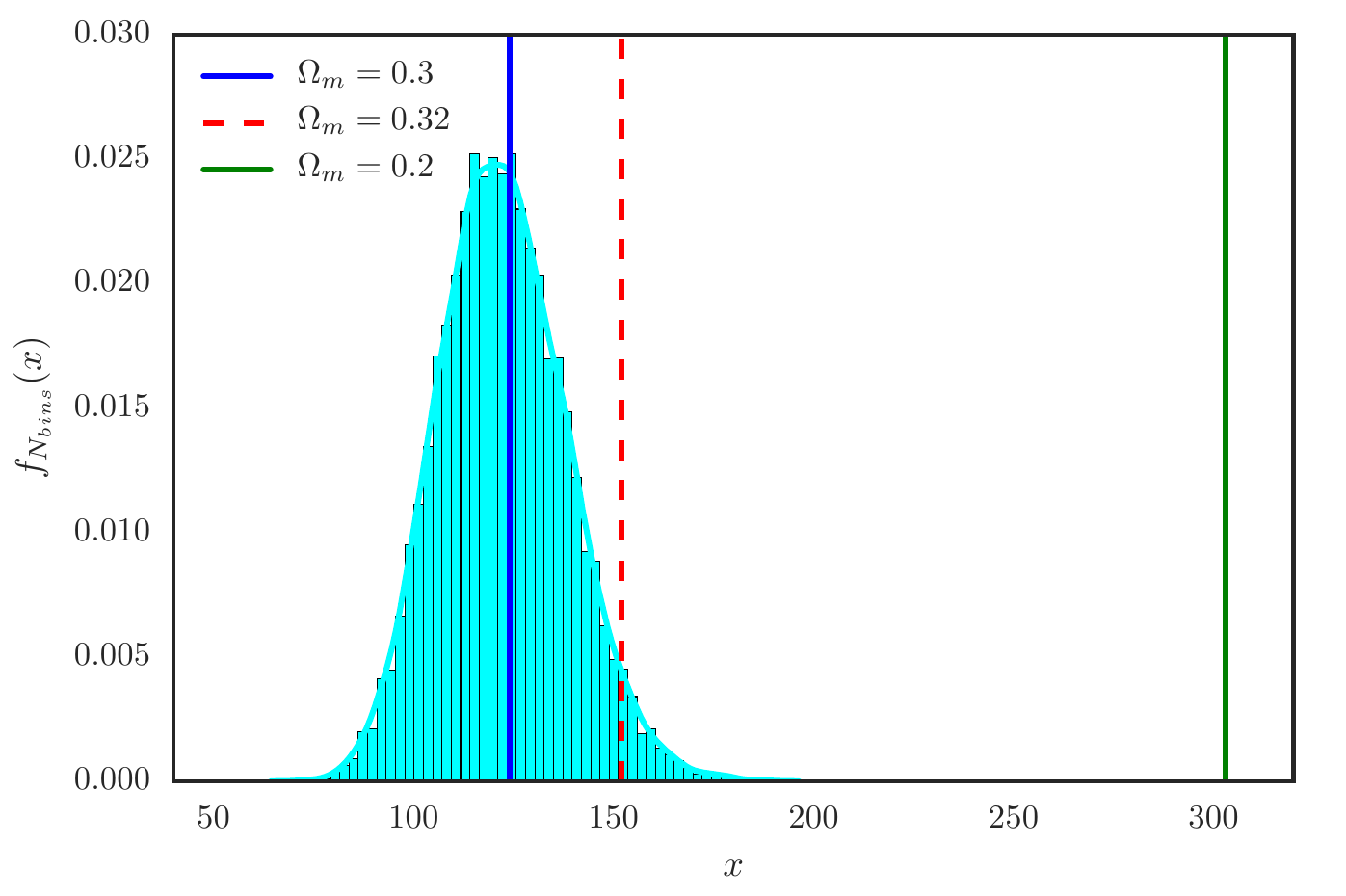}
\caption{The $\chi^2$ distribution with $N_{bins}$ degrees of freedom where $x=\chi^2/dof$, (cyan).  The values we obtain for Eq. \ref{eq:chi2} in   trial cosmologies of
$\Omega_m^* = 0.3,0.32,0.2$ are plotted as blue solid, red dashed and green solid vertical lines respectively.}
\label{fig:chi2}
\end{center}
\end{figure}

As noted above the test statistic for the Light Curve metric is distributed according to a $\chi^2$ distribution with $N_{\rm bins}$ degrees of freedom ($dof$) so in our simple example with 4 different trial cosmologies, we can ask: what is the probability we obtain this value, $x=\chi^2/dof$,  given the degrees of freedom.
 The relevant $\chi^2$ distribution is plotted in Fig. \ref{fig:chi2} as a cyan histogram together with the values we obtain for the distance metric in trial cosmologies of
$\Omega_m^* = 0.3$ (blue solid), $0.32$ (red dashed), and $0.2$ (green solid). 
The fact
that the blue line lies at a high probability with respect the the cyan distribution (p-value ~0.025) compared to the red dashed line (p-value~0.005) means we would favor the cosmology represented by the blue line over the one represented by the red.
Overall this statistic shows a clear trend to favor the simulation which has $\Omega_m^*=0.3$ over a simulation with $\Omega_m^* = 0.32$ or 0.2. Note the  $\chi^2/dof$ for $\Omega_m^*=0.4$ is 510 and so does not appear in the range plotted.

\section{Example wall clock times for running the {\it superABC} sampler} \label{app:C}

Ultimately the run times for any ABC sampler will depend most heavily on the forward model simulation which is where the majority of the time is spent in the algorithm. However there are several settings which can decrease time to convergence and make the sampling algorithm more efficient.
Throughout this work we use the {\it superABC} sampler with an adaptive tolerance threshold based on the 75$^{\rm th}$ quartile of the distances in the previous iteration and a weighted covariance matrix estimation in order to perturb the particles at each iteration \citep[see][for more details on these settings]{2016arXiv160807606J}.
Using the Tripp metric (Section \ref{sec:tripp}) without systematic effects and 100 particles on 96 compute nodes the sampler takes $\sim24$ hours to complete 10 iterations, at which time the particles are clearly sampling from the posterior distribution (burn-in can be clearly seen in the trace plots of individual particles we have examined).
After this the sampler slows considerably as the threshold level decreases and the posterior distribution is being sampled from. In this work we run the particles until the uncertainty on the $1\sigma$ errors is $\sim$1\%. Typically this will take another 4-5 days of running with the same number of compute nodes. Including Planck priors speeds up the run time by approximately a factor of 2 for the initial steps and a factor of $20\%$ after $\sim10$ iterations.
The MCMC sampler was run with 100 chains with 4000 steps in each ($\sim $40000 samples from the posterior distribution). This took $\sim$24-36 hours on 16 cores depending on whether or not systematics where included or the choice of prior.

\section{Constraining supernovae standardization parameters} \label{app:D}

To demonstrate that the Tripp metric, discussed in Section \ref{sec:tripp} can constrain the SN standardization parameters, $\alpha$ and $\beta$, correctly to within similar accuracy to a maximum likelihood approach, we present results from the {\it superABC} sampler using the data set described in Section \ref{sec:data}.
In this example all parameters apart from $\alpha$ and $\beta$ are fixed at the true values.

\begin{figure}
\begin{center}
\includegraphics[height=1.5in,width=3.5in]{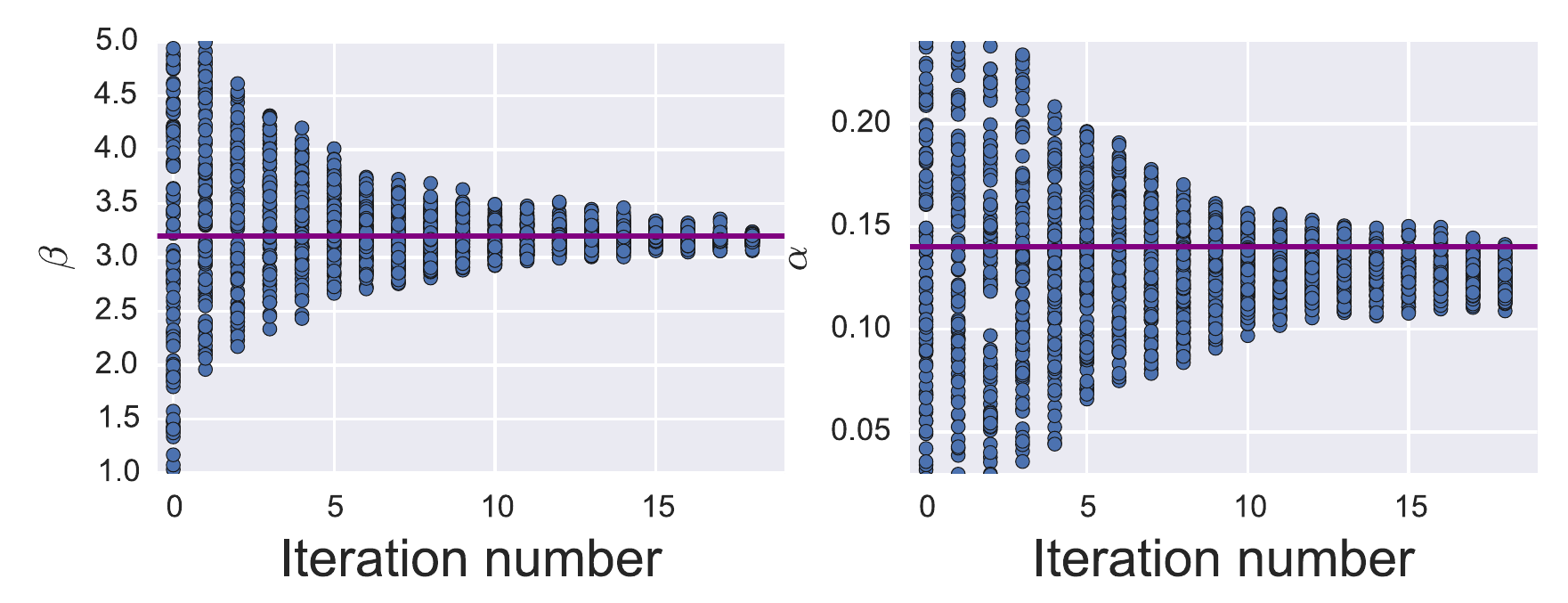}
\caption{The accepted parameter values for $\beta$ ($\alpha$) at each iteration are shown in the left (right) panel. Each particle in the {\it superABC} sampler is shown as a blue dot. The `true'' value of each parameter is shown as a purple solid line in each panel.}
\label{fig:2_param_1}
\end{center}
\end{figure}

\begin{figure}
\begin{center}
\includegraphics[height=3in,width=3.2in]{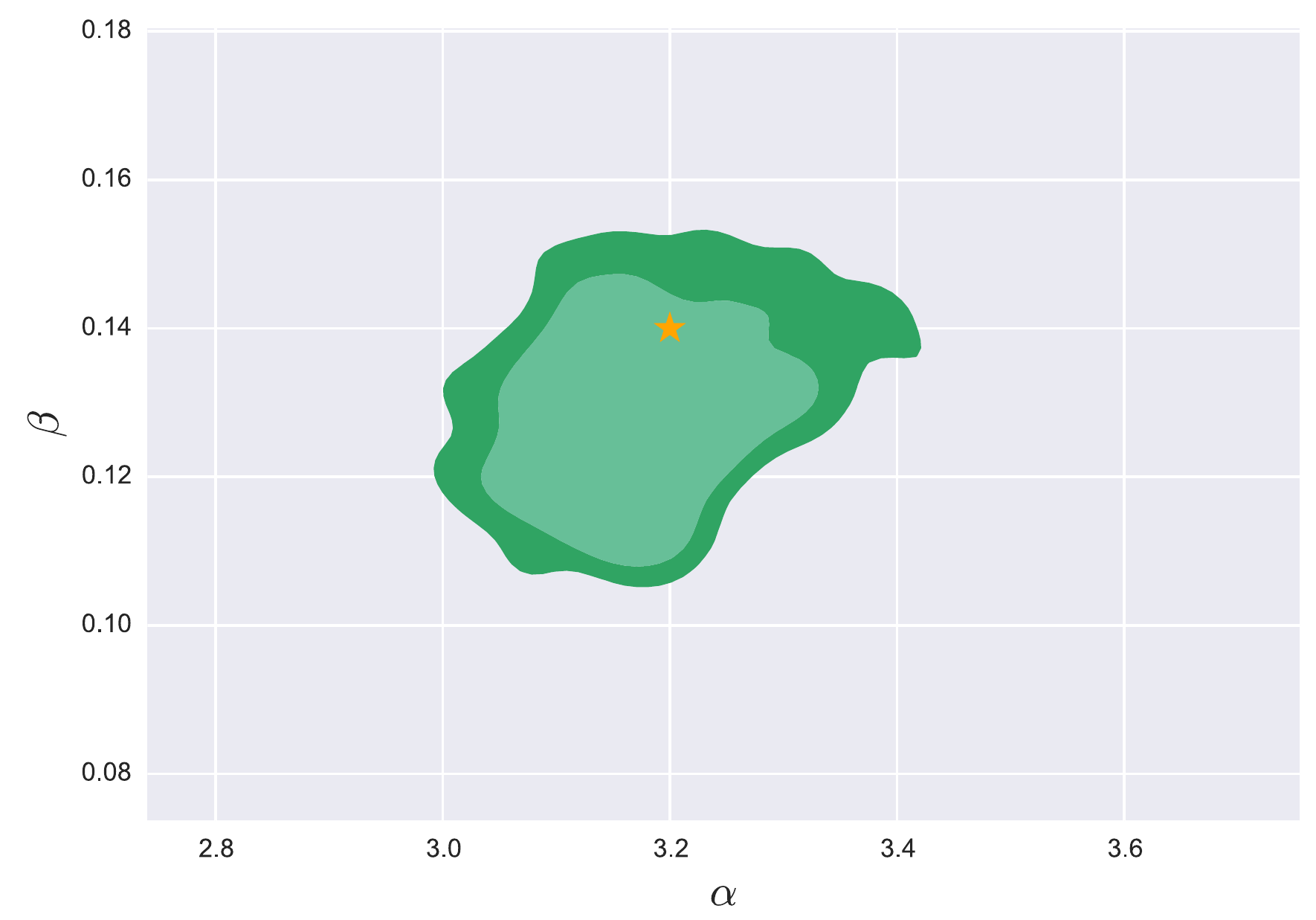}
\caption{The 1 and 2$\sigma$ contours for $\alpha$ and $\beta$ using the Tripp metric and 100 particles for 19 iterations.
The filled green contours show constraints obtained by varying only these 2 parameters in the Tripp metric described in Section \ref{sec:tripp}.
The ``true" parameters of the data are represent by the yellow star.}
\label{fig:2_param_2}
\end{center}
\end{figure}

In Fig.\ \ref{fig:2_param_1} we show the accepted parameter values for $\beta$ ($\alpha$) at each iteration in the left (right) panel. Each particle in the {\it superABC} sampler is shown as a blue dot and the `true' value of each parameter is shown as a purple solid line. At iteration 0 the particles start out widely dispersed throughout the uniform prior range. At subsequent iterations the particles clearly converge towards the correct values of the parameters as the accepted tolerance threshold decreases.
In Fig.\ \ref{fig:2_param_2} the 1 and 2$\sigma$ contours for $\alpha$ and $\beta$ after 19 iterations of 100 particles are shown as
light and dark green filled contours respectively. The true parameters of the data are represent by the yellow star.
The 1 $\sigma$ constraints on the standardization parameters are:
$\alpha = 0.1285 \pm 0.011$ and $\beta = 3.176 \pm 0.060$. 
Using the SALT2mu fitter on the same data the constraints are $\alpha = 0.15755 \pm 0.0078$ and  $\beta=3.159 \pm 0.069$.
Note that a direct comparison between the results of the SALT2mu maximum likelihood technique 
and a Bayesian sampler such as {\it superABC} is not strictly valid given the differences in the two methods, the interpretation of 1$\sigma$ errors in each and the effect of priors. We give the SALT2mu constraints here as readers familiar with  this technique may wish to make a rough comparison. 

\section{Some features of the superABC sampler}\label{app:E}
There are several features of {\it superABC} which are designed to optimize the sampling procedure: 
\begin{itemize}
\item the sampler can be run in parallel, either using Python's mpi4py\footnote{https://pypi.python.org/pypi/mpi4py} or multiprocessing\footnote{https://docs.python.org/2/library/multiprocessing.html}, such that each particle runs its own simulation concurrently with other particles in one iteration,
\item the code creates a hash table (a python dictionary where the key is the unique id for the simulated light curve) for fast lookup of simulation outputs needed in the distance metric,
\item a python wrapper, {\it rootpy}\footnote{http://www.rootpy.org/}, is available to read {\sc ROOT}\footnote{https://root.cern.ch/} outputs if available which can decreases i/o time substantially compared to reading ascii,
\item only particle parameter values, weights and distances are saved at every iteration such that the simulation output files can be overwritten during sampling in order to save on i/o and disk space.
\end{itemize} 
The full end user options are documented on the wiki at https://github.com/EliseJ/superabc

\bibliographystyle{mn2e}
\bibliography{thebibliography}

\end{document}